\documentclass[usenatbib]{mnras}

\usepackage{newtxtext,newtxmath}
\usepackage{braket}
\usepackage{caption}
\usepackage{subcaption}
\usepackage{booktabs}
\usepackage{multirow}
\usepackage{tabularx}

\usepackage[T1]{fontenc}

\DeclareRobustCommand{\VAN}[3]{#2}
\let\VANthebibliography\thebibliography
\def\thebibliography{\DeclareRobustCommand{\VAN}[3]{##3}\VANthebibliography}

\usepackage{graphicx}	
\usepackage{amsmath}	
\usepackage[dvipsnames]{xcolor}
\usepackage{ulem}
\usepackage{cancel}

\newcommand{\msun}{~\rm M_{\large \odot}}
\newcommand{\lsim}{\mathrel{\hbox{\rlap{\lower.55ex\hbox{$\sim$}} \kern-.3em\raise.4ex\hbox{$<$}}}}

\title[EMRIs triggered by MBHBs]{Extreme Mass Ratio Inspirals triggered by Massive Black Hole Binaries: from Relativistic Dynamics to Cosmological Rates.}

\author[G. Mazzolari et al.]{Giovanni Mazzolari,$^{1,2,3}$\thanks{E-mail: giovanni.mazzolari@inaf.it}
Matteo Bonetti,$^{4,5,6}$\thanks{E-mail: matteo.bonetti@unimib.it}
Alberto Sesana$^{4,5,6}$,
Riccardo M. Colombo$^{4}$,
Massimo Dotti$^{4,5,6}$,
\newauthor
Giuseppe Lodato$^{3}$,
David Izquierdo-Villalba.$^{4,5}$
\\
$^{1}$Department of	Physics	and	Astronomy, Alma Mater Studiorum Universit\'a di Bologna, via Gobetti 93/2, I-40129 Bologna, Italy\\
$^{2}$INAF - Astrophysics and Space	Science	Observatory	of Bologna, via Gobetti 93/3, I-40129 Bologna, Italy\\
$^{3}$ Dipartimento di Fisica "Aldo Pontremoli", Università degli Studi di Milano, via G. Celoria 16, I-20133 Milano, Italy\\
$^{4}$Dipartimento di Fisica ``G. Occhialini'', Universit\`a degli Studi di Milano-Bicocca, Piazza della Scienza 3, I-20126 Milano, Italy\\
$^{5}$INFN, Sezione di Milano-Bicocca, Piazza della Scienza 3, I-20126 Milano, Italy\\
$^{6}$INAF - Osservatorio Astronomico di Brera, via Brera 20, 20121 Milano, Italy\\
}

\date{Accepted 2022 August 3. Received 2022 July 25; in original form 2022 April 12}

\pubyear{2015}

\begin{document}
\label{firstpage}
\pagerange{\pageref{firstpage}--\pageref{lastpage}}
\maketitle

\begin{abstract}
Extreme mass ratio inspirals (EMRIs) are compact binary systems characterized by a mass-ratio $q=m/M$ in the range $~10^{-9}-10^{-4}$ and represent primary gravitational wave (GW) sources for the forthcoming Laser Interferometer Space Antenna (LISA). 
While their standard formation channel involves relaxation processes deflecting compact objects on very low angular momentum orbits around the central massive black hole, a number of alternative formation channels has been proposed, including binary tidal break-up, migration in accretion disks and secular and chaotic dynamics around a massive black hole binary (MBHB).
In this work, we take an extensive closer look at this latter scenario, investigating how EMRIs can be triggered by a MBHBs, formed in the aftermath of galaxy mergers.
By employing a suite of relativistic three-body simulations, we evaluate the efficiency of EMRI formation for different parameters of the MBHB, assessing the importance of both secular and chaotic dynamics. By modelling the distribution of compact objects in galaxy nuclei, we estimate the resulting EMRI formation rate, finding that EMRI are produced in a sharp burst, with peak rates that are 10-100 times higher than the standard two-body relaxation channel, lasting for 10$^6$--10$^8$ years. By coupling our results with an estimate of the cosmic MBHB merger rate, we finally forecast that LISA could observe ${\cal O}(10)$ EMRIs per year formed by this channel.
\end{abstract}



\section{Introduction}

Galactic nuclei are of great interest for understanding a wide variety of phenomena ranging from large-scale galaxy evolution to relativistic dynamics.
They are characterised by typical stellar densities that in the central parsec can reach (or even exceed) the remarkable level of $10^6\msun$ pc$^{-3}$, and they usually host a massive black hole \citep[MBH, see e.g.][and references therein]{Merrit2013,Kormendy_2013}, that can reveal its presence via a number of violent phenomena sourced by its strong gravitational field. Perhaps the most common and best studied of which is efficient accretion of ionized plasma, powering quasars and spectacular relativistic jets \citep[e.g.][]{1979ApJ...232...34B,1984ARA&A..22..471R}.

Besides accreting gas, MBHs also interact with stars and compact objects (COs). In particular, two-body encounters within the dense environment can scatter objects close to the MBH horizon, giving rise to a wide variety of phenomena driven by extreme, relativistic dynamics, such as stellar tidal disruptions \citep{1988Natur.333..523R}, hypervelocity star ejections \citep{1988Natur.331..687H} and relativistic captures of COs \citep{2001CQGra..18.4067H,Seoane2018}. To the last class do extreme mass ratio inspirals (EMRIs) and direct plunges (DPs) belong, the main difference between the two being the specific dynamics of the COs after gravitational capture by the MBH. If the CO is scattered on an eccentric orbit with pericentre smaller than the last stable orbit (which is $4GM/c^2$ for an almost parabolic orbit around a non-rotating MBH of mass $M$), it will directly fall onto the central MBH without significant loss of gravitational radiation, thus resulting in a DP. Conversely, if the pericentre is $\gtrsim 4GM/c^2$, the CO is captured on a bound orbit, progressively releasing its orbital energy in form of gravitational waves (GWs) mostly emitted as bursts at each pericentre passage. This process results in a slow inspiral, lasting millions of orbits. Since COs are generally neutron stars (NSs) or stellar black holes (BHs) of mass $m=1-10 M_\odot$ and the central MBH has mass $10^4M_\odot<M<10^9M_\odot$, the mass-ratio of the binary is extremely small, $q=m/M=10^{-9}-10^{-4}$, which is why they are called EMRIs.

Since they are expected to emit GWs in the milli-Hz frequency band, EMRI detection is among the primary goals of the future Laser Interferometer Space Antenna \citep[LISA,][]{2017arXiv170200786A}.
Indeed, these sources are anticipated to be extraordinary tools for astrophysics, fundamental physics and cosmology. Because of the extreme ratio between the two body masses, the CO essentially acts as a test mass orbiting the central MBH, thus offering a unique possibility of mapping its spacetime and testing putative deviations from the Kerr solution \citep[e.g.][]{2007PhRvD..75d2003B,2010arXiv1009.1402A,2019NatAs...3..447H}. EMRIs' complex gravitational waveforms will allow a precise estimation of the source parameters including masses, the MBH spin, the MBH quadrupole mass moment, sky location and luminosity distance \citep{Barak_cutler_2004}. The latter could be used to infer the galaxy host via statistical method, allowing to estimate cosmological parameters even in absence of any electromagnetic counterparts to the GW source \citep{2008PhRvD..77d3512M,2021MNRAS.508.4512L}. The EMRI mass and eccentricity distribution will provide precious information about the dynamical processes shaping galactic nuclei and about the low mass end of the MBH mass function, which is currently poorly constrained \citep{2010PhRvD..81j4014G,2019ApJ...883L..18G}.

On a dynamical standpoint, forming 
EMRIs requires putting a CO on a very tight (semi-major axis $a$ between $10^{-2} - 10^{-4}$ pc), low--angular-momentum (eccentricity $e>0.999$) orbit around the central MBH \citep{Seoane2007}. This can occur as a result of different dynamical mechanisms.

In the standard channel, extensively covered in literature \citep{Alexander,2011PhRvD..84d4024M,Seoane2013,EMRI_LISA,Seoane2018}, EMRIs are formed around a single MBH at the center of a dense stellar nucleus as a consequence of two-body encounters (i.e. two-body relaxation). Because of the high densities in the central regions, COs, and in particular stellar BHs, gravitationally interact with other bodies, continuously changing their orbital parameters. A repeated sequence of two-body interactions can set the CO on an almost radial orbit entering the GW dominated regime, thus forming an EMRI.
A second possibility is EMRI formation via binary separation. When a binary system composed by at least one stellar BH is scattered close to the central MBH, the tidal forces can separate the two bodies and the stellar BH can be delivered to the MBH on an EMRI orbit \citep{Miller_2005}.
Alternatively, \citet{Pan_2021} studied EMRI formation in the context of COs migration within the innermost part of the accretion disk of the MBH, a scenario initially proposed by \citet{2007MNRAS.374..515L}. In particular, stellar BHs and stars on inclined orbits are first captured by the accretion disk, and then subsequently migrate towards the MBH under the influence of density wave generation and head wind. This channel is not well constrained yet, but according to \citet{Pan_2021} it might significantly contribute to the total rate.

This paper is dedicated to another dynamical channel, that can lead to a burst of EMRI formation due to the presence of a massive black hole binary (MBHB). These binaries are expected to form in the aftermath of galaxy mergers, as first theorized by \cite{1980Natur.287..307B}. The presence of a second MBH delivered to the galaxy center exerts strong perturbations onto the nuclear distribution of stars and COs, possibly triggering a number of interesting phenomena. For example, it has been shown that the presence of a MBHB can trigger a burst of TDEs, \citep{Chen2008,Chen2009,2011ApJ...729...13C,Chen_2012,Chen2013,10.1093/mnras/stv1031,Ricarte2016} as a result of a combination of secular effects such as Lidov-Kozai oscillations \citep[LK,][]{Lidov,Kozai,Naoz2016} and chaotic interactions. The very same processes could also promote the formation of EMRIs. This formation channel has received little attention in the literature so far, with the exception of the seminal work of \cite{BodeWegg}. While completing this work, we also became aware of the study by \cite{Naoz2022}. We will discuss similarities and differences with both studies.

We study the detailed relativistic dynamics of a triple system formed by a stellar BH and a MBHB. We characterise how the MBHB's presence triggers EMRI formation and compute EMRI rates as a function of the properties of the system.
The final goal is to constrain the cosmological EMRI formation rate sustainable by this channel and estimate the number of events that can be detected by LISA. The manuscript is organized as follows.
In Section~\ref{sec:EMRI formation} we discuss the theoretical background of EMRI formation around MBHs and MBHBs. In Section~\ref{sec:methods} we introduce the features of the code used for the simulations and the initial conditions of the physical set-up. We then present the methods used to extract the relevant information from the simulations: the EMRI formation rate for single MBHBs (Section~\ref{sec:sis rate 1}), the cosmological EMRI formation rate from MBHBs (Section~\ref{sec: cosmological rate 1}), and the expected MBHB-triggered EMRI detection rates with LISA (Section~\ref{sec:Lisa rate 1}). In Section~\ref{sec:results} we present and discuss the results obtained. Finally, in Section~\ref{sec:conclusions} we present our final considerations, including a detailed comparison with previous works, and an extensive discussion of caveats, limitations and future plans. 

\section{Theoretical framework} \label{sec:EMRI formation}

\subsection{Standard formation scenario}

We start by reviewing the basics of the standard capture mechanism relying on two-body relaxation. Such process is a direct consequence of the granularity of galactic systems, in which the single components (e.g. stars) are subjected to random kicks due to the interactions with the whole particle ensemble. The accumulation of small perturbations changes the energy and angular momentum of stars over a relaxation time \citep{Merrit2013}:

\begin{equation} \label{t_merrit}
     t_{\rm rlx} = 1.2 \times 10^{11} \Biggl( \frac{\sigma}{100 \rm km s^{-1}} \Biggr)^{7.47} \rm yr,  
\end{equation}
which is the typical timescale needed by a star to experience a change in its velocity magnitude $\delta v\approx v$. In equation~\eqref{t_merrit}, $\sigma$ is the velocity dispersion of stars, that in turn can be related to the mass of the central MBH through the empirically observed $M-\sigma$ relation, \citep{Gebhardt,Ferrarese}:
\begin{equation}\label{eq:M-sigma}
    \sigma(M_\bullet) = 200 \biggl(\frac{M_\bullet}{3.09 \times 10^8 M_{\rm \odot}}\biggr)^{1/4.38} \rm km/s .
\end{equation}
It follows that for a galaxy with a central MBH with $M_\bullet=10^6 M_{\odot}$ the typical time needed by a galaxy bulge to be relaxed is of the order of $t_{\rm rlx}= 1.5$ Gyr. 

One of the main consequences of relaxation is mass segregation that leads more massive objects to migrate towards the inner regions and the lighter ones towards the external regions. As reported in \cite{Preto_seoane} and in \cite{EMRI_LISA}, the timescale over which mass segregation happens is a fraction of the relaxation timescale and approximately reads:
\begin{equation}
    t_{\rm sgr}= (0.1-0.25)\ t_{\rm rlx}.
\end{equation}
We can therefore expect mass segregated cusps to be common around MBHs with $M_\bullet\lesssim 10^7M_{\odot}$,
with densities reaching $\rho\sim 10^6-10^8 \rm M_{\odot}  pc^{-3}$. 
Interactions within the cusp continuously change the energy and angular momentum of the orbiting bodies, scattering some of them on very eccentric orbits, with a pericentre grazing the MBH event horizon. Depending on the nature of the deflected object and the pericentre distance we can have three main different outcomes:

\begin{enumerate}
    \item {\it direct plunges (DP)}: the deflected object is directly captured by the MBH after being scattered on an orbit that will bring the object directly into the event horizon.
    \item {\it tidal disruption events (TDEs)}: the deflected object is disrupted by the tidal field exerted by the MBH. Note that this outcome is possible only for extended objects (e.g. main sequence stars).
    \item {\it EMRIs}: the object is deflected onto a very eccentric orbit grazing the MBH last stable orbit, such that GW emission is sufficiently strong to decouple it from stellar perturbations. In order to survive the tidal field on such an extreme orbit, the object must be compact, i.e. a stellar-mass BH or a neutron star.
\end{enumerate}

Here we focus on EMRIs. As already mentioned, successful EMRI formation depends on the synergy between GW-emission and relaxation. On the one hand, GW-emission extracts energy and angular momentum from the CO's orbit. At the lowest radiative Post-Newtonian (PN) order the (osculating) orbital elements evolve as \citep{Peters}
\begin{align}
\begin{split}
\label{eqn:GW_ae}
    &\frac{da}{dt}= -\frac{64}{5}\frac{G^3\mu M^2}{c^5 a^3 } F(e) \\
    &\frac{de}{dt}= -\frac{304}{15}\frac{G^3\mu M^2}{c^5 a^4 (1-e^2)^{-5/2}}\biggl(e+\frac{121}{304}e^3\biggr),
\end{split}
\end{align}
with the function $F(e)$ given by
\begin{equation}
    F(e) = (1-e^2)^{-7/2}\biggl(1+ \frac{73}{24}e^2+\frac{37}{96}e^4\biggr).
\end{equation}
The very steep dependence of $F(e)$ on the eccentricity determines that the time-scale of the evolution $T_{\rm GW} (a,e)$ decreases by orders of magnitude as $e$ tends to unity, therefore increasing the probability of decoupling the system MBH-BH from the surrounding environment. On the other hand, in fact,
two body relaxation can randomly scatter the CO orbit in and out of an EMRI trajectory.
Those perturbations occur over a 
time-scale $\approx (1-e)t_{\rm tlx}$ \citep{Seoane2007}, and a proper EMRI forms when the following condition is satisfied
\begin{equation}\label{eq:EMRI_c}
    T_{\rm GW}(a,e)\leq (1-e)t_{\rm rlx},
\end{equation}
meaning that after the first deflection, the GW inspiral time is shorter than the timescale of other orbital perturbations.
The estimated rate at which COs (mainly stellar BHs) become EMRIs via this standard scenario is between $\sim 10^{-6}$ and $10^{-8}$ $\rm yr^{-1}$ \citep{Seoane2007}, considering a MBH of $10^6M_{\rm \odot}$. As reported in \cite{EMRIrate} the EMRI formation rate scales with the mass of the central MBH as:
\begin{equation} 
    \mathcal{R}\propto M^{-\beta}_\bullet \qquad \beta\in [1/4, 1],
\end{equation}
meaning that the heavier the MBH is, the smaller is the EMRI rate.\\
Moreover, considering uncertainties connected with the poorly constrained cosmic number density and spatial distribution of MBHs in the mass range $M_\bullet\in[10^4,10^7]M_{\rm \odot}$ (the sweet spot of LISA sources), the predicted number of events that will be observed by LISA spans from a few to a few thousands per year \citep{EMRI_LISA}.
Given this quite large uncertainty range it is worth exploring other possible formation scenarios.

\subsection{EMRI formation in three-body systems}

We consider an EMRI formation scenario in which, rather than relying on two-body relaxation, an additional incoming MBH ($M_2$) is responsible for deflecting COs ($m_3$) toward the central MBH ($M_1$). We therefore consider a three-body system which is hierarchical in nature, i.e. it features, in its starting configuration, a primary MBH surrounded by a CO cusp and a secondary MBH approaching from larger scales. A hierarchical triplet is characterised by the presence of two well separated binaries: the inner binary formed by $M_1-m_3$ and an outer binary comprised by the centre of mass of the inner one and $M_2$. Such kind of systems can be analysed in the framework of the secular theory \citep[see e.g.][and references therein]{Ford2000,Holman1997}, in which, through dedicated averaging techniques, only variations on a timescale much longer than the orbital periods are considered. 
A special feature of hierarchical triplets, that has been extensively explored within the secular theory, is the LK mechanism \citep{Kozai,Lidov}. Those authors, by perturbatively expanding the equations of motion of the three-body system in terms of the small ratio between the inner and outer semi-major axes ($\alpha=a_{\rm in}/a_{\rm out} \ll 1$), found that the inner and outer binaries can exchange angular momentum. Such exchanges happen above a certain inclination threshold ($\approx 39^\circ$) and trigger oscillations between the eccentricity of the inner binary and the relative inclinations of the orbital planes of the two binaries.
More specifically, at the lowest order of expansion in $\alpha$ (quadrupole level), assuming $m_3$ a test mass and considering initial circular orbits, it is possible to show that the quantity\footnote{Assuming that the whole system has the total angular momentum along the $z$-direction, this quantity is nothing else than the $z$-component of the inner orbit angular momentum, which at the lowest level of pertubative expansion is exactly conserved \citep[see e.g.][]{Naoz}.}
\begin{equation}\label{eq:LK1}
    \sqrt{1-e^2_{\rm in}}\cos{(\iota_{\rm rel})},
\end{equation}
is a constant of motion, implying that every variation of the inner eccentricity is reflected in an opposite variation of the relative inclination and viceversa. It can be shown that the LK oscillations are most effective for nearly polar orbits, i.e. $\iota_{\rm rel}\sim 90^{\circ}$, for which during the oscillation $e_{\rm in}$ can reach values close to unity. Further considering higher-order terms in the expansion (octupole, exadecapole, etc.), a richer and more complex phenomenology arises \citep[see e.g.][]{Naoz2016,Will2017,Lim2020}, possibly leading to even larger eccentricities that, when dealing with compact objects, imply an extremely powerful emission of GWs and a remarkable speed-up of the GW inspiral.

However, the specific physical system that we are going to study in this work presents a few additional complications. Firstly, the outer binary has an evolving semi-major axis as the secondary MBH ($M_2$), due to the interaction with the stellar environment in which it resides, gets closer and closer to the inner binary. This means that at a certain point in the evolution the hierarchy between the inner and outer binaries may cease to exist, invalidating the assumptions behind the secular approximation and demanding different techniques to deal with the chaotic dynamics. The separation between the two regimes depends on the characteristics of the three-body system and takes place at a characteristic outer separation \citep{Mardling}
\begin{equation}\label{eq:caotic}
    a_{\rm chaos}\approx \frac{3.3}{1-e_{\rm  \rm out}}\biggl[\frac{2}{3}\biggl(1+\frac{M_2}{M_1 + m_3}\biggr) \frac{1+e_{\rm  \rm out}}{\sqrt{1-e_{\rm  \rm out}}}\biggr]^{2/5} \biggl(1-\frac{0.3 \iota_{\rm  \rm rel}}{\pi}\biggr)a_{\rm  \rm in}.
\end{equation}
Therefore for $a_{\rm out} \leq a_{\rm chaos}$ the system cannot be analysed within the framework of the secular theory, but instead demands to solve the full set of equations of motion.

Secondly, given the relativistic nature of the objects involved, GR effects cannot be safely neglected. Apart from the dissipative effect of GWs, another crucial phenomenon that has to be considered is the relativistic precession at pericentre. Indeed, it has been found that precession tends to destroy the coherency that leads to the development of LK oscillations \citep{Holman1997,Blaes2002,Miller2002,2018PASA...35...17B,Lim2020}, therefore when the precession timescale is shorter than the LK one, the eccentricity growth can be severely suppressed. 

\begin{figure}
    \centering
    \includegraphics[width=\columnwidth]{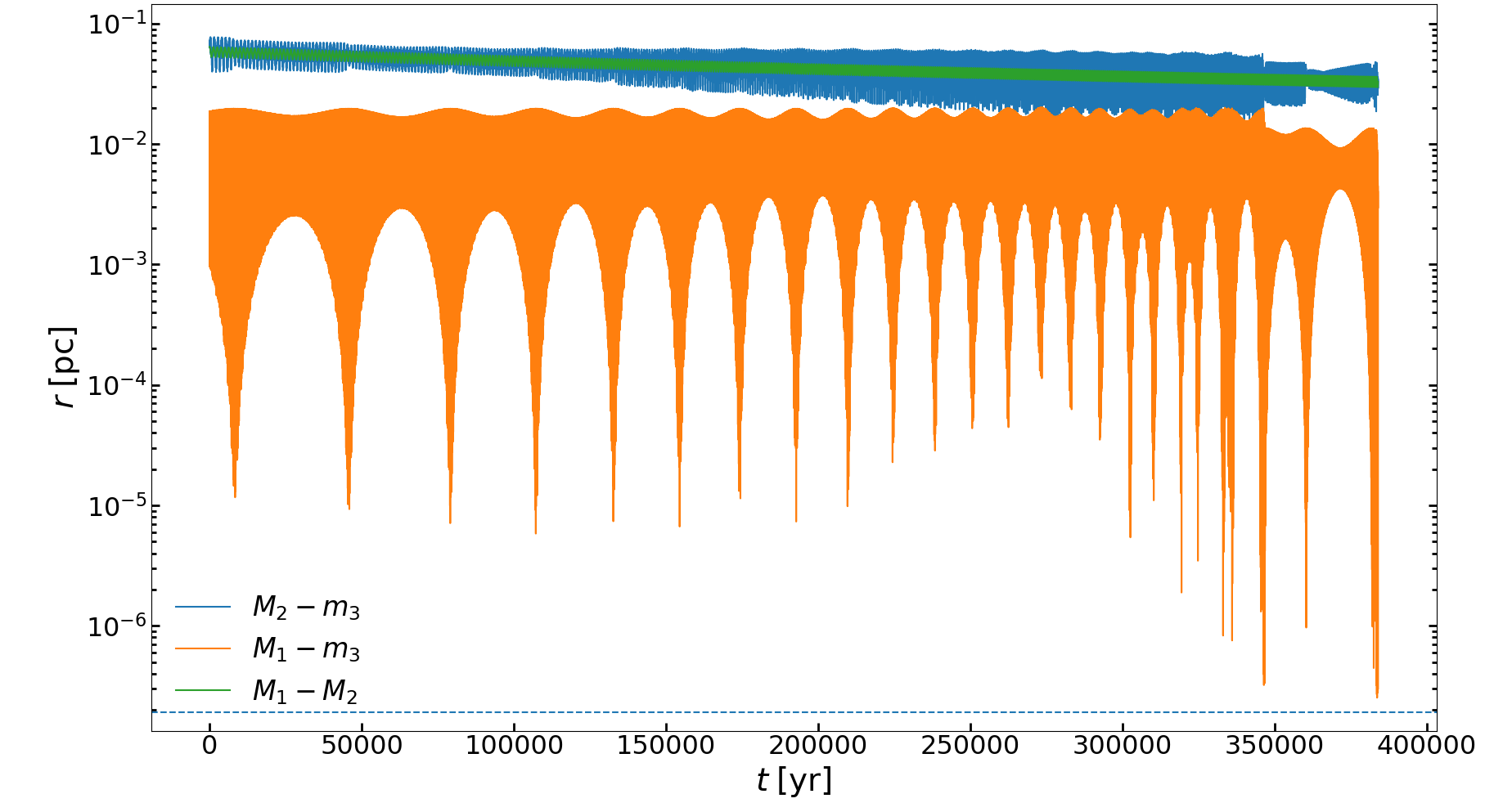}
    \caption{Time evolution of the relative distances between each pair of the three involved bodies (color code as labelled). In this specific simulation we consider $M_1=10^6 M_{\rm \odot}$, $q=0.1$, $e_{\rm out}=0.1$ and the mass of the stellar BH is set to $m_3=10 M_{\rm \odot}$. As time passes the perturbation exerted by $M_2$ accumulates and LK oscillations start as visible by the large spread (i.e. higher eccentricity) in the relative separation between $M_1$ and $m_3$ (orange line). Also note how the separation between $M_1$ and $M_2$ (green line) is decreasing because of the action of stellar hardening.}
    \label{fig:Kozai}
\end{figure}

Fig.~\ref{fig:Kozai} shows an example of the evolution of a three-body system, in which the inner binary with an extreme mass-ratio is perturbed by an incoming MBH. From the figure it is possible to see the effect of LK oscillations on a stellar BH that has been identified as an EMRI at the end of one of our simulations. Looking at the relative distance between $M_1$ and $m_3$ (represented in orange) it is possible to see that the position of the pericentre $r_p=a(1-e)$ is affected by a series of large periodic oscillations, due to the eccentricity oscillations associated to LK. While the position of $r_p$ varies of a factor between $\sim 10^{2}-10^{4}$ during an oscillation, the position of the apocentre of the orbit, $r_a=a(1+e)$, is not much affected by the LK mechanism during the evolution. From the plot it is also possible to observe that the period of the oscillation in eccentricity is much longer than the orbital period of both the binaries, confirming the fact that LK is a secular effect.

In order to capture at best the phenomenology of the extreme system under study we therefore numerically integrate the full set of three-body equations of motion including GR corrections up to the 2.5 PN order, that we are going to discuss in the next section.


\section{Methods} \label{sec:methods}

\subsection{Computational setup} \label{Description of the code}
To study EMRI formation triggered by MBHBs, we employ the three-body integrator originally developed in \cite{Bonetti1}. Specifically, the code leverages on a C++ implementation of the Burlish-Stoer (BS) algorithm based on the Modified Midpoint Algorithm and the Richardson extrapolation \citep{Richardson1911,Bulirsch1966,NR}. The code numerically solves the Hamiltonian equations of motion of the three-body system in the centre of mass frame. Further to standard Newtonian dynamics, GR corrections are introduced according to the perturbative Post-Newtonian scheme up to the 2.5 PN order\footnote{Note that with the PN scheme we can avoid implementing a pseudo-Newtonian potential or pausing the simulation in order to account for GW losses, as done in \cite{BodeWegg}.}
\begin{equation}\label{PN}
    H = H_{\rm 0} + \frac{1}{c^2} H_{\rm 1} + \frac{1}{c^4} H_{\rm 2} + \frac{1}{c^5}H_{\rm 2.5} + O\biggl(\frac{1}{c^6}\biggr).
\end{equation}
In the above Hamiltonian, terms up to 2 PN are conservative and essentially introduce relativistic pericentre precession, that e.g. for the inner binary reads
\begin{equation}\label{eq:GR_prec}
    \delta\omega_{ \rm GR} = \frac{6\pi G (M_1 + m_3)}{a_{ \rm in}(1-e_{ \rm in})^2c^2} + \frac{3 \pi G^2 (18 + e_{ \rm in}^2)(M_1 + m_3)^2}{2 a_{ \rm in}^2(1-e_{ \rm in}^2)^2 c^4},
\end{equation}
while the 2.5 PN term is the lowest radiative order and accounts for GW energy dissipation. Since we are considering only non-spinning black holes the 0.5 PN and the 1.5 PN terms, capturing the spin-orbit coupling of the binary, are not taken into account. We defer this improvement to a future, updated version of the code.

Further to the three-body dynamics, the code allows to consider the effect of the stellar distribution that is expected to surround MBHBs \citep[see][for details]{Bonetti1}. Specifically, the stellar background is modelled through a spherically symmetric mass distribution following an Hernquist profile \citep{1990ApJ...356..359H}. 

The stellar potential provides an additional conservative force the motion of the three compact objects is subjected to, with the most noticeable effect given by an additional precession (opposite to PN precession) of the form
\begin{equation}\label{eq:NW_prec}
    \delta\omega_{ \rm star}=-2\pi  \frac{\sqrt{1-e^2}}{1 + \sqrt{1-e^2}}\frac{M_{*}(<r)}{M_1},
\end{equation}
where
\begin{equation}\label{mass_encl}
    M_{*}(<r) = M_{\rm bulge}\biggl(\frac{r}{r+\lambda}\biggr)^{2},
\end{equation}
is the stellar mass enclosed in the orbit of a body at separation $r$ from the stellar potential's origin. 
Following \citet{Sesana_Khan}, we anchor the properties of the stellar distribution to the mass of the MBHB through the scaling relations
\begin{align} \label{eq:MBHB_features}
    &M_{\rm bulge} = 4,8 \times 10^{8}\biggl(\frac{M_1 + M_2}{10^6 M_{\rm \odot}}\biggr)^{1/1.16} \ \rm M_{\rm \odot}, \\
    &\lambda= 65\biggl(\frac{M_1 + M_2}{10^6 M_{\rm \odot}}\biggr)^{0.596/1.16}\ \rm pc.
    \label{eq:R_MBHB}
\end{align}
For the present study we decided to fix the centre of the stellar potential on the primary MBH ($M_1$). This choice is motivated by the fact that within the sphere of influence we expect that the dynamics is dominated by the MBH itself, therefore any perturbation on $M_1$ (e.g. that exerted by the secondary MBH) is also reflected on the stellar cusp around $M_1$. We checked that this specific choice does not affect our findings, in particular we verified that the statistics of forming EMRIs does not noticeably depend on the chosen origin of the stellar profile (see Appendix~\ref{sec:appA}).

Finally, the stellar environment also extracts energy and angular momentum from the MBHB, a process known as stellar hardening \citep{Quinlan1996, Sesana2006}. This phenomenon is due to the collective effect of three-body interactions between a bound MBHB and single passing stars, which after a series of chaotic interactions get generally ejected with positive energy extracted from the binary. Since we do not feature a live galactic nucleus sampled with particles, this process is captured by our code through the addition of a dissipative force acting on MBHs and tuned to reproduce the orbital averaged evolution predicted for the MBHB semi-major axis and eccentricity \citep[see e.g.][]{Quinlan1996,Sesana2006,Bonetti1}, i.e.
\begin{align}\label{hard}
    \Dot{a} &= -a^{2}\frac{G\rho H}{\sigma}\\
    \Dot{e} &= a\frac{G\rho H K}{\sigma}.
    \label{hard2}
\end{align}
Here $H$ and $K$ are respectively the dimensionless hardening rate and the eccentricity growth rate  \citep[taken from][]{Sesana2006}, while $\rho$ and $\sigma$ are the density and velocity dispersion at the influence radius of the MBHB \citep{Sesana_Khan}. We operationally define the influence radius $R_{\rm inf}$ as the radius enclosing twice the MBHB's mass in stars, which for an Hernquist profile reads 
\begin{equation}
    R_{\rm inf} = \frac{(2M_{\rm MBHB})^{1/2}}{M_{\rm bulge}^{1/2}-(2M_{\rm MBHB})^{1/2}} \lambda,
\end{equation}
that through the scaling relations \eqref{eq:MBHB_features} and \eqref{eq:R_MBHB} becomes
\begin{equation}
    R_{\rm inf}= 4.5 \left(\frac{M_{\rm MBHB}}{10^6 M_{\rm \odot}}\right)^{\frac{1.352}{2.32}} \frac{0.935}{1-0.065\left(\frac{M_{\rm MBHB}}{10^6 M_{\rm \odot}}\right)^{\frac{0.16}{2.32}}} \ \rm pc.
\end{equation}

\subsection{Initial conditions} \label{setup}

We initialise bound three-body systems comprising an inner binary, formed by the primary MBH, $M_1$, orbited by a stellar-mass BH $m_3$, and an outer binary, formed by the intruder MBH, $M_2$, and the centre of mass of the inner binary. In practice, given the large mass difference between $M_1$ and $m_3$ the centre of mass coincides with the position of $M_1$. All bodies are subjected to the additional gravitational acceleration of the stellar distribution.

\begin{figure}
    \centering
    \includegraphics[width=\columnwidth]{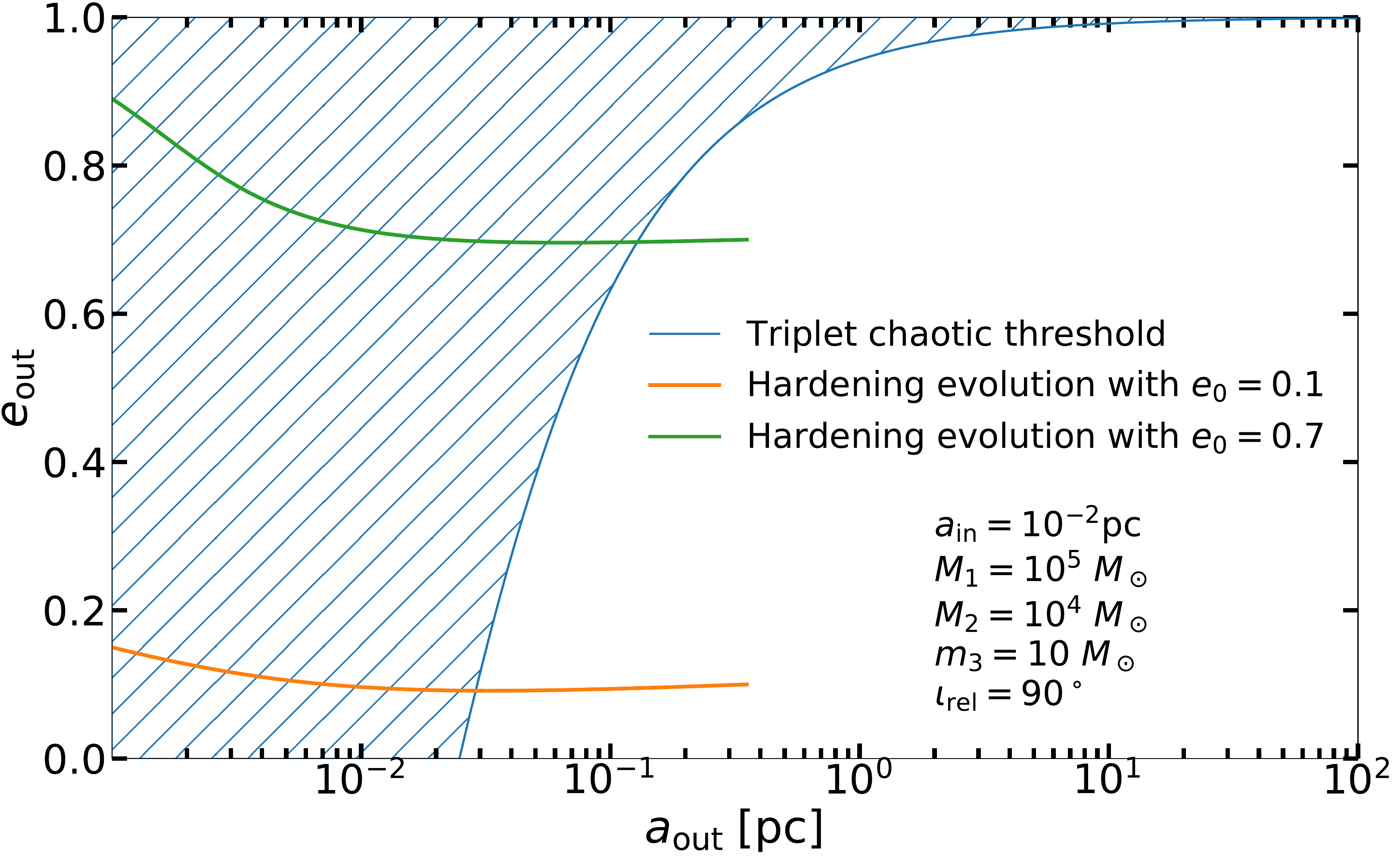}
    \caption{Comparison between the triplet chaotic threshold and two hardening evolution tracks (corresponding to two different values of $e_{\rm out}$) in the $e_{\rm out}-a_{\rm out}$ plane. The three tracks are computed considering the values of the parameters reported in figure. When one given hardening track intersects the chaotic threshold the triplet enters the chaotic regime (hatched zone).}
    \label{fig:a_chaos}
\end{figure}

We choose a typical value for the stellar BH mass, i.e. $m_3 = 10 M_{ \rm \odot}$, while we select three values for the mass of the primary MBH, $M_1=3\times10^5 M_{ \rm \odot}$, $10^6 M_{ \rm \odot}$, $3\times10^6 M_{ \rm \odot}$, which produce EMRIs emitting GWs in the mHz range that will be surveyed by LISA.
We then set the mass of the secondary MBH, $M_2$, considering four different values of mass-ratio $q=M_2 / M_1=0.1$, $0.03$, $0.01$, $0.003$. We avoid mass-ratios close to unity since in this limit it is expected that the strong torque exerted by the MBHB ejects most of the stellar mass BHs, therefore suppressing EMRI formation \citep[an effect already seen by][while studying TDE formation in an analogous setup]{2011ApJ...729...13C}. 
Finally, we considered two different initial eccentricities selected at MBHB formation, namely $e_{\rm out}=0.1, 0.7$. 
The above choice of parameters gives 24 different combinations and for each of them we then initialise 20000 simulations varying the orbital properties as follows:
\begin{itemize}
    \item the orbital angles of the inner binary, specifically the argument of pericentre ($\omega_{\rm in}$), the longitude of the ascending node ($\Omega_{\rm in}$) and the relative inclination ($\iota_{\rm rel}$) are assigned assuming isotropy, with $\omega_{\rm in}$ and $\Omega_{\rm in}$ lying within $[0,2\pi]$ and $\cos \iota_{\rm}$ in $[-1,1]$. The outer binary is assumed to lie in the $x-y$ plane with $\omega_{\rm out}$ and $\Omega_{\rm out}$ set to zero for simplicity.
    
    \item the inner binary eccentricity $e_{\rm in}$ is drawn from a thermal distribution, i.e. $f(e_{\rm in}) = 2e_{\rm in}$ \citep{thermalecc}. Such distribution is expected in dense environments where stellar-mass objects receive random kicks due to their mutual interaction.
    
    \item the inner binary's semi-major axis is drawn from a log-flat distribution with limits given by
    \begin{equation}\label{range}
        \biggl(\frac{M_1}{10^6 M_{ \rm \odot}}\biggr)^{\alpha}\times 10^{-3} {\rm pc} < a_{ \rm in} < \biggl(\frac{M_1}{10^6 M_{ \rm \odot}}\biggr)^{\alpha} \times 10^{-1} {\rm pc},
    \end{equation}
    where the dependence on the primary mass has been empirically set to $\alpha=0.8$, so that the upper limit of the simulated range coincides with the maximum semi-major axis that can result in the formation of an EMRI, derived from Eq. ~\eqref{eq:EMRI_c}.
    
    \item Finally, in order to assign the outer semi-major axis and eccentricity we first need to ensure the stability of the three-body system, i.e. we need $a_{\rm out} > a_{\rm chaos}$ (see Eq.~\eqref{eq:caotic}). In principle, one can easily ensure stability by setting an arbitrarily large $a_{\rm out}$, but this would imply a waste of computing time as the secondary sinks to the center exerting a negligible influence on the inner  binary. We found a balance between initial stability and computational efficiency by setting
    \begin{equation}\label{eq:initial_a_out}
        a_{\rm out} = 2 a_{\rm chaos}(a_{\rm in},e_{\rm out}).
    \end{equation}
    However, since $a_{\rm chaos}$ depends not only on $a_{\rm in}$ but also on $e_{\rm out}$ (that also evolves for stellar hardening), $a_{\rm out}$ is not readily assigned. Thus, we first compute the evolutionary track of $(a_{\rm out}, e_{\rm out})$ for a given MBHB and we then select the value of $a_{\rm out}$ satisfying Eq.~\eqref{eq:initial_a_out} together with the corresponding outer eccentricity. This procedure is visualized in Fig~\ref{fig:a_chaos}.\\

    Before proceeding, we comment here on the chosen initial separation of the binary. We show that, although the adopted value is arbitrary, it is in fact sufficient to capture the relevant dynamical evolution of the system, and therefore to appropriately estimate the EMRI rate. If it is true that LK oscillations can operate also for $a_{\rm out} > 2 a_{\rm chaos}$ and therefore can possibly produce EMRIs also for larger initial $a_{\rm out}$, we also expect that those EMRIs will be subdominant in number because of a combination of effects. Since the stellar hardening timescale scales as $T_{\rm hard}\propto a_{\rm out}^{-1}$, the time spent to shrink the binary from e.g. $5 a_{\rm chaos}$ to $2 a_{\rm chaos}$ is almost half the time spent from $2 a_{\rm chaos}$ to $a_{\rm chaos}$, implying much less time for LK to develop. Moreover, since the LK timescale itself scales with a quite steep power of the outer semi-major axis, i.e. $T_{\rm LK}\propto a_{\rm out}^3$, it is unlikely that successful LK oscillations could strongly affect the inner binary and produce an EMRI in the shorter $T_{\rm hard}$ available. It should also be noted that the LK oscillation timescale is in competition with the relativistic periastron precession timescale of the inner binary, which means that longer oscillations are more easily quenched, further suppressing EMRI formation.
    Finally, we note that despite our systems start into a stable hierarchical configuration, the action of the stellar hardening can drive the triplets into an unstable configuration where $a_{\rm out} < a_{\rm chaos}$, where strong chaotic encounter dominates the dynamical evolution of the system and is the primary channel for EMRI formation.
\end{itemize}

\subsubsection{Stellar hardening efficiency}
\label{sec:hardening}

In some of the simulations, especially those with a tight inner binary, the computational time may be significant since the three-body integrator has to adapt the time-step to follow the motion of the inner binary on a fraction of its (small) orbital period. This means that, for quite small $a_{\rm in}$, before $M_2$ could exert a noticeable effect the integrator has to perform a fairly large number of steps.
Therefore, in order to save computational time and so accelerate the simulations (and for the computation of the MBHB hardening only), the density at the influence radius of the MBHB entering in the hardening equations \eqref{hard} and \eqref{hard2} is empirically tuned according to the following expression
\begin{equation}\label{eq:rhoeff}
    \frac{\rho_{ \rm eff}}{\rho} = \frac{2}{5} \biggl[\biggl(\frac{M_1}{10^6 M_{ \rm \odot}}\biggr)^{\alpha}\biggl(\frac{1pc}{a_{ \rm in,0}}\biggr)\biggr]^{\log 5}.
\end{equation}
This artificial increase translates into a direct speed-up of the hardening phase and triggers a quicker interaction between the inner and outer binaries when the initial $a_{\rm in}$ is smaller. We verified on a sub-sample of simulations that this choice does not impact the number of EMRIs or DPs and their overall properties.

\subsection{Simulation stoppage conditions} \label{sec:stoppage}

The integration proceeds until one of the following four outcomes occurs:
\begin{itemize}
    \item[1.] {\it Ejection}: when one of the three bodies is far enough from the other two, the system is no more a bound triplet. The only cases of ejections reported in our simulations are the ejections of $m_3$, which is much lighter than the MBHs. The ejection is flagged when two conditions are satisfied, i.e. the distance between $M_1$ and $m_3$ exceeds the scale radius of the Hernquist profile $\lambda$ and the total energy of the ejected body is positive.
    
    \item[2.] {\it Swap}: this happens when the secondary MBH substitutes $m_3$ in the inner binary, but without ejecting it. In this case the two original binaries have swapped. Specifically, such event is identified when:
    \begin{equation}
        r_{\rm p,in} = a_{\rm in}(1-e_{\rm in}) > a_{\rm out}(1+e_{\rm out})= r_{\rm a,out}.
    \end{equation}
    This condition means that the pericenter of the inner binary becomes larger than the apocenter of the initial outer binary, and so the position of $m_3$ and $M_2$ are reversed. In practice $m_3$ has been pushed on a weakly bound orbit around the MBHB, which is extremely unlikely to result in either an EMRI or a DP.
    
    \item[3.] {\it Direct plunge (DP)}: when the relative separation between $m_3$ and either $M_1$ or $M_2$ is less than 
    \begin{equation}
        d_{\rm i3} \leq 4 \frac{G(M_{\rm i} + m_{\rm 3})}{c^2},
    \end{equation}
    approximately corresponding to the last stable orbit for a massive particle on an highly eccentric orbit around a Schwarzschild BH, then no EMRI formation takes place and $m_3$ is considered directly captured by the MBH.
    
    \item[4.] {\it EMRI}: a successful EMRI forms when the following conditions are met
    \begin{equation} \label{EMRI_conditions}
    \begin{cases}
      T_{\rm GW}(a_{\rm in}, e_{\rm in}) \leq (1-e_{\rm in})t_{\rm rlx} \\
      a_{\rm in}(1+e_{\rm in})\leq \frac{1}{4} u(\Tilde{q}) a_{\rm out}(1-e_{\rm out})
    \end{cases}
    \end{equation}
    The first condition ensures that the system has decoupled form the stellar environment, i.e. the GW-driven inspiral dominates over two-body relaxation. In our system, however, because of the presence of an additional MBH, we need also to ensure that $M_2$ does not interfere with the GW inspiral. The second condition therefore checks that $m_3$ (in particular its apocentre) lies within the influence sphere of $M_1$, i.e. within its Roche Lobe. The size of the primary Roche Lobe (in units of outer binary semi-major axis) can be computed considering the approximate solution by \citet{Eggleton1983}
    \begin{equation}
         u(\Tilde{q})=\frac{0.49\Tilde{q}^{2/3}}{0.6\Tilde{q}^{2/3}+\ln(1+\Tilde{q})^{1/3}}, \quad \Tilde{q}=\frac{M_1}{M_2},
    \end{equation}
    which depends only on the mass-ratio. The above solution is strictly valid only for circular binaries, we thus compute the Roche lobe of the primary as if the outer binary was circular with radius equal to its pericentre $r_p=a_{\rm out}(1-e_{\rm out})$, obtaining $r_{\rm Roche}=u(\Tilde{q}) a_{\rm out}(1-e_{\rm out})$. Moreover, we require that the apocentre of the inner binary has to be smaller than $1/4$ of the dimension of the new Roche Lobe for a successful capture.\footnote{We checked that our findings do not sensibly depend on the arbitrary threshold of $1/4$. In fact, changing this number from 1 to 1/4, the number of formed EMRIs converged quite fast: we witnessed a noticeable difference going from 1 to 1/2, while from 1/2 to 1/4 such number is not sensibly affected.
    }
\end{itemize}

\subsection{EMRI rate computation} \label{sec:computation of emri rate}

In order to estimate the cosmological merger rate of EMRIs triggered by MBHBs, the outcomes of the three-body simulations are coupled to a  semi-analytical framework as we now illustrate. We first describe how the obtained number of EMRIs can be rescaled for different initial CO cusp profiles, then we detail how EMRI rates derived for individual systems are coupled with the cosmological MBHB merger rate, as obtained from semi-analytical galaxy  formation models.

\subsubsection{Rates from three-body simulations: cuspy profiles} \label{sec:sis rate 1}
As said, as a consequence of mass segregation, it is expected that MBHs are surrounded by a steep cusp of compact objects, that can be approximately described to follow a singular isothermal sphere (SIS) profile:

\begin{equation}
    \rho_{ \rm BH}= \frac{\sigma^2}{2\pi G r^2}.
\end{equation}
with $\sigma$ denoting the normalisation of the profile and physically representing the velocity dispersion.

Given the SIS mass density, the number of stellar mass BH distributed in an interval of semi-major axis between $a$ and $a + da$ is
\begin{equation}
 \frac{d\mathcal{N}_{ \rm BH,SIS}}{da}=\int_0^{2\pi}\int_0^{\pi}\frac{\rho_{ \rm BH}(a)}{m_{ \rm BH}} a^2 \sin{\theta}d\theta d\phi = \frac{2\sigma^2}{m_{ \rm BH}G},
\end{equation} 
that after a straightforward integration gives
\begin{equation}
    \mathcal{N}_{ \rm BH,SIS}(a)= \frac{2\sigma^2}{m_{ \rm BH}G} a .
    \label{eq:nbhsis}
\end{equation}
If we now divide the variability range of the BH's semi-major axis in 20 log bins equally spaced, with separations $x_j$, $x_{j+1}$, the number of BHs in the $j$-th bin will be:
\begin{equation}
    \mathcal{N}_{ \rm BH,SIS}(j) = \frac{2\sigma^2}{m_{ \rm BH}G} (10^{x_{j+1}}-10^{x_j}).
\end{equation}
On the contrary, since we took a log-flat distribution for $a_{\rm in}$ as initial condition in our simulations and we sampled 20000 total BHs, their number in every log bin is constant and given by $N_{ \rm BH,sim}=1000$. Using the following proportion:
\begin{equation}
    \mathcal{N}_{ \rm EMRI, SIS}(j) : \mathcal{N}_{ \rm BH, SIS}(j) = N_{ \rm EMRI, sim}(j) : N_{ \rm BH,sim}
\end{equation}
and defining the weights:
\begin{equation}\label{pesi}
    w_j = \frac{\mathcal{N}_{ \rm BH, SIS}(j)}{N_{ \rm BH,sim}},
\end{equation}
we finally obtained the expected number of EMRIs in a SIS-like cusp starting from those detected in simulations:
\begin{equation}\label{N_emri}
    \mathcal{N}_{ \rm EMRIs,SIS} = \sum_j N_{ \rm EMRI, sim}(j) \ w_j.
\end{equation}

Next, in order to assess the rate, we need the time that EMRIs take to form and evolve. This time will be the time taken by the MBHB to shrink plus the time that the GW inspiral takes after we stop the three-body simulations, i.e.
\begin{equation}\label{T_tot}
    T_{ \rm tot}= T_{ \rm hard} + T_{ \rm GW}.
\end{equation}
We evaluate $T_{ \rm hard}$ as the time needed to shrink the MBHB from the formation, $a_{ \rm out,0}$, down to $a_{ \rm out,f}$, where the simulation is stopped, i.e.
\begin{equation}
   T_{ \rm hard} = \frac{\sigma_{ \rm inf}}{G \rho_{ \rm inf}}\int_{ \rm a_{ \rm out,f}}^{a_{ \rm out,0}} \frac{da}{a^2 H(a)},
\end{equation}
where $\rho_{\rm inf}$ is taken as the actual value of the density (and not the enhanced one used to boost the MBHB hardening) and $H(a)$ is a numerical parameter encoding the efficiency of stellar scattering \citep{Sesana2006}.
For the evaluation of $T_{ \rm GW}$, namely the time the $m_3$ inspiral lasts (from the EMRI identification until the plunge), we instead employ the expression of the time to coalescence found by \citet{Peters}, i.e.
\begin{align}
    T_{ \rm GW}(a,e) &= \frac{12}{19} \frac{c_0^{4}}{\beta} \nonumber\\ 
    &\times \int_{0}^{e} \frac{e^{29/19} [1+(121/304)e^2]^{1181/2299}}{(1-e^2)^{3/2}}de, 
\end{align}
where 
\begin{align}
    c_0&=a\ e^{-12/19}(1-e^2)\biggl(1+\frac{121}{304}e^2\biggr)^{-870/2299},\\ 
    \beta&=\frac{64}{5}\frac{G^3M_1m_3(M_1+m_3)}{c^5}. 
\end{align}

Finally the rates for a SIS-like BH cusp are computed dividing $\mathcal{N}_{\rm EMRI, SIS}$ in Eq.~\eqref{N_emri}, by the averaged time of Eq.~\eqref{T_tot}, $\Braket{T_{\rm tot}}$ i.e.

\begin{equation}
\label{eq:emrirate}
    \frac{d\mathcal{N}_{\rm EMRI, SIS}}{dt} \approx \frac{\mathcal{N}_{\rm EMRI, SIS}}{\Braket{T_{\rm tot}}}.
\end{equation}

The result of this procedure will be presented in Tab.~\ref{tab:EMRI_rates} and discussed in Sec.~\ref{sec: sis rate 2}.

By using Eq.~\eqref{eq:nbhsis} to compute $\mathcal{N}_{\rm BH}$, we are implicitly assuming that the central cusp is composed by BHs only. Although BHs are expected to be outnumbered by stars by a factor of $10^3$, segregation processes are expected to drive the BHs to the center, pushing stars outwards. This was demonstrated by \cite{2006ApJ...649...91F}, who found that, for a MW-like system, BHs dominate the density distribution out to $\approx 0.3$ parsec, consistent with the inner binary's semi-major axis range sampled by our simulations (cf. Eq.~\eqref{range}).

\subsubsection{MBHB merger rate} \label{sec: cosmological rate 1}

As the presence of a MBHB is a necessary condition for the EMRI formation process studied here, a meaningful assessment of the cosmological EMRI merger rate requires a model for the formation rate of MBHBs across the Universe. 
We extract this information from the semi-analytical cosmological model \textit{L-Galaxies} \citep{L-gal,Izquierdo2020}. 

\textit{L-Galaxies} exploits the merger tree of the Millennium II dark matter only simulation,\footnote{The Millennium II follows the cosmic evolution of $2160^3$ DM particles of mass $6.9 \times 10^6 M_{\rm \odot}$ contained in a periodic comoving cube of 100 Mpc/\textit{h} on a side ,  using cosmological parameters: $\Omega_{\rm m} = 0.315$, $\Omega_b = 0.045$, $\Omega_{\rm \Lambda} = 0.685$, and $h = 0.673 \ \rm  km\ s^{-1}Mpc^{-1}$.} representing the skeleton on top of which the semi-analytic model runs. 
When \textit{L-Galaxies} recognizes the formation of a DM halo more massive than $\approx 10^8 M_{\rm \odot}$, a proportional amount of baryonic matter ($\sim 15\%$ of the halo mass) and a seed MBH of mass $M_{\rm seed}=10^{3} M_{\rm \odot}$ are placed at the center of the halo. Then \textit{L-Galaxies} follows the evolution of every resolved halo (and its central MBH) considering baryon cooling, star formation and stellar structure development (disk and bulges), MBH accretion and  merger between galaxies (and MBHs). Specifically, the version of \textit{L-Galaxies} that we used did not include the dynamical evolution of MBHBs following galaxy mergers, i.e. the two MBHs coalesce instantaneously as the two host galaxies merge. Including dynamical delays might change the outcome redshift distribution of MBHBs in non-trivial ways -- an effect that we will explore in future work.

We also discarded coalescences involving MBHs below $2\times 10^3M_\odot$ (i.e. still close to the seed mass). This choice is motivated by the fact that the semi-analytic model does not have the resolution to carefully track low-mass mergers. Moreover, as we will see, LISA is not sensitive to EMRIs around MBHs with  $M\lesssim 10^4M_\odot$.

From \textit{L-Galaxies} we extracted a total number of $N=391.691$ mergers and we constructed the cosmic merger rate as follows. First, we built the differential comoving number density ${d^3 n}/({d\log M \ d\log q \ dz })$  by binning events in a 3-D grid of $M_1$, $q$ and $z$\footnote{The grid features 60 logarithmically equally spaced bins for $M_1$ and $q$, with $M_1 \in [2\times 10^3, 10^{10}] M_{\rm \odot}$ and $q \in [10^{-6}, 1]$, while $z$ is binned in 60 linearly equally spaced bin with $z \in [0,8]$.} and dividing by the simulation comoving volume $V_c$.
Then the observed merger rate at a given redshift is obtained by integrating over the comoving volume shell through the relation
\begin{align}\label{eq:m_rate_2}
     \frac{d^4 N}{d\log M \ d\log q \ dz \ dt_{\rm 0}} &= \frac{d^3  n}{d\log M \ d\log q \ dz } \cdot \frac{dV_{\rm c}}{dz} \cdot \frac{dz}{dt_{\rm r}} \cdot \frac{dt_{\rm r}}{dt_{\rm 0}} \\
     & = \frac{d^3  n}{d\log M \ d\log q \ dz } \cdot 4\pi c D_M^2,
\end{align}
where $D_M$ is the comoving distance and $t_r = t_0/(1+z)$ is the relation between rest-frame and observed time.

The cosmological EMRI rate is finally computed simply by multiplying $\mathcal{N}_{\rm EMRI, SIS}$ in Eq.~\eqref{N_emri} by the MBHB merger rate given by Eq.~\eqref{eq:m_rate_2}, to obtain
\begin{equation}\label{eq:rate_tot}
   \frac{d^4 N_{\rm EMRI, tot}}{d\log M \ d\log q \ dz \ dt_{\rm 0}}=\mathcal{N}_{\rm EMRI, SIS} \cdot \frac{d^3  n}{d\log M \ d\log q \ dz } \cdot 4\pi c D_M^2
\end{equation}
Numerically, the result is achieved by applying a bi-linear interpolation between the two $M-q$ grids over which the number of EMRIs and the merger rate are defined. In particular, we multiplied the result of Eq.~\eqref{eq:m_rate_2} by $\mathcal{N}_{\rm EMRI,SIS}$, using the central values of $M$ and $q$ of each bin, and limiting the $M$ and $q$ ranges to those used for the EMRIs simulations, namely $M\in[3\times 10^5, 3\times 10^6] M_{\rm \odot}$ and $q\in [0.003, 0.1]$. 
The results are presented and discussed in Section~\ref{sec: cosmological rate 2}.

\subsubsection{LISA detection rates} \label{sec:Lisa rate 1}

In order to compute the expected LISA detection rate for EMRIs, we employ the simple analytic kludge (AK) waveform model developed by \citet{Barak_cutler_2004} and implemented in \citet{Bonetti2020}. The AK model is a leading order Fourier-domain waveform that leverages on the Newtonian fluxes worked out in \cite{Peters} to evolve the orbital elements of binary systems, i.e. orbital frequency (related to the semi-major axis) and eccentricity. The AK model then computes an inclination-polarisation--averaged characteristic strain that relates to how GW power is distributed among several harmonics of the orbital motion. The waveform is then used to compute the signal-to-noise ratio (SNR) assuming the sky-averaged LISA sensitivity curve \citep{2021arXiv210801167B}, where we fix an EMRI detection threshold to an $\rm SNR = 20$ \citep{Barak_cutler_2004,Colpi_Sesana_2017}.
The expected detection rates are presented in Section~\ref{sec:Lisa rate 2}.

\section{Results} \label{sec:results}

\subsection{EMRI Dynamical formation}

\begin{table}
    \setlength\extrarowheight{3pt}
    \centering
    \begin{tabular}{cc|ccccc}
        \toprule
        \multicolumn{2}{c|}{$M_1=3\times 10^5 M_{\odot}$} & \multirow{2}{*}{EMRIs} & \multirow{2}{*}{DPs} & \multirow{2}{*}{Swap} & \multirow{2}{*}{Ejections} & \multirow{2}{*}{Unresolved}  \\
        $q$ & \multicolumn{1}{c|}{$e_{\text{out},0}$} &  &  &  &  &    \\
        \midrule
        \multirow{2}{*}{$0.1$} & \multicolumn{1}{c|}{$0.1$} & $1044$ & $2466$ & $7568$ & $8921$ & $1$  \\
                               & \multicolumn{1}{c|}{$0.7$} & $294$ & $157$ & $6937$ & $12613$ & $0$ \\
        \hline
        \multirow{2}{*}{$0.03$} & \multicolumn{1}{c|}{$0.1$} & $1185$ & $2279$ & $4923$ & $11608$ & $5$ \\
                                & \multicolumn{1}{c|}{$0.7$} & $379$ & $211$ & $5432$ & $13978$ & $0$  \\
        \hline
        \multirow{2}{*}{$0.01$} & \multicolumn{1}{c|}{$0.1$} & $1001$ & $1180$ & $4797$ & $13012$ & $10$ \\
                                & \multicolumn{1}{c|}{$0.7$} & $612$ & $399$ & $5237$ & $13751$ & $1$ \\
        \hline
        \multirow{2}{*}{$0.003$} & \multicolumn{1}{c|}{$0.1$} & $799$ & $584$ & $5958$ & $12616$ & $43$ \\
                                 & \multicolumn{1}{c|}{$0.7$} & $1094$ & $558$ & $5590$ & $12750$ & $8$ \\
        \bottomrule
    \end{tabular}
    \begin{tabular}{cc|ccccc}
        \toprule
        \multicolumn{2}{c|}{$M_1= 10^6 M_{\odot}\ \ \ \ \ \ $} & \multirow{2}{*}{EMRIs} & \multirow{2}{*}{DPs} & \multirow{2}{*}{Swap} & \multirow{2}{*}{Ejections} & \multirow{2}{*}{Unresolved}  \\
        $q$ & \multicolumn{1}{c|}{$e_{\text{out},0}$} &  &  &  &  &    \\
        \midrule
        \multirow{2}{*}{$0.1$} & \multicolumn{1}{c|}{$0.1$} & $272$ & $2391$ & $8031$ & $9324$ & $0$  \\
                               & \multicolumn{1}{c|}{$0.7$} & $140$ & $206$ & $7016$ & $12638$ & $0$ \\
        \hline
        \multirow{2}{*}{$0.03$} & \multicolumn{1}{c|}{$0.1$} & $247$ & $2194$ & $5379$ & $12180$ & $0$ \\
                                & \multicolumn{1}{c|}{$0.7$} & $201$ & $283$ & $5441$ & $14075$ & $0$  \\
        \hline
        \multirow{2}{*}{$0.01$} & \multicolumn{1}{c|}{$0.1$} & $265$ & $1231$ & $5020$ & $13480$ & $4$ \\
                                & \multicolumn{1}{c|}{$0.7$} & $348$ & $518$ & $5338$ & $13796$ & $0$ \\
        \hline
        \multirow{2}{*}{$0.003$} & \multicolumn{1}{c|}{$0.1$} & $386$ & $683$ & $6315$ & $12606$ & $10$ \\
                                 & \multicolumn{1}{c|}{$0.7$} & $615$ & $737$ & $6011$ & $12631$ & $6$ \\
        \bottomrule
    \end{tabular}
    \begin{tabular}{cc|ccccc}
        \toprule
        \multicolumn{2}{c|}{$M_1=3\times 10^6 M_{\odot}$} & \multirow{2}{*}{EMRIs} & \multirow{2}{*}{DPs} & \multirow{2}{*}{Swap} & \multirow{2}{*}{Ejections} & \multirow{2}{*}{Unresolved}  \\
        $q$ & \multicolumn{1}{c|}{$e_{\text{out},0}$} &  &  &  &  &    \\
        \midrule
        \multirow{2}{*}{$0.1$} & \multicolumn{1}{c|}{$0.1$} & $50$ & $2115$ & $8515$ & $9320$ & $0$  \\
                               & \multicolumn{1}{c|}{$0.7$} & $71$ & $282$ & $6837$ & $12810$ & $0$ \\
        \hline
        \multirow{2}{*}{$0.03$} & \multicolumn{1}{c|}{$0.1$} & $53$ & $1851$ & $5704$ & $12391$ & $1$ \\
                                & \multicolumn{1}{c|}{$0.7$} & $85$ & $335$ & $5563$ & $14017$ & $0$  \\
        \hline
        \multirow{2}{*}{$0.01$} & \multicolumn{1}{c|}{$0.1$} & $60$ & $1059$ & $5498$ & $13383$ & $0$ \\
                                & \multicolumn{1}{c|}{$0.7$} & $181$ & $602$ & $5391$ & $13825$ & $1$ \\
        \hline
        \multirow{2}{*}{$0.003$} & \multicolumn{1}{c|}{$0.1$} & $140$ & $844$ & $6770$ & $12245$ & $1$ \\
                                 & \multicolumn{1}{c|}{$0.7$} & $351$ & $922$ & $6452$ & $12275$ & $0$ \\
        \bottomrule
    \end{tabular}
    \caption{Final outcomes of the simulations divided according to the mass of the primary, the mass-ratio and the initial outer eccentricity.
    The last column contains the number of simulations that kept reaching the integration time limit of $150$ minutes after three restarts, which are discarded.}
    \label{tab:risultati_sim}
\end{table}

\begin{figure}
\centering
 	\includegraphics[width=\columnwidth]{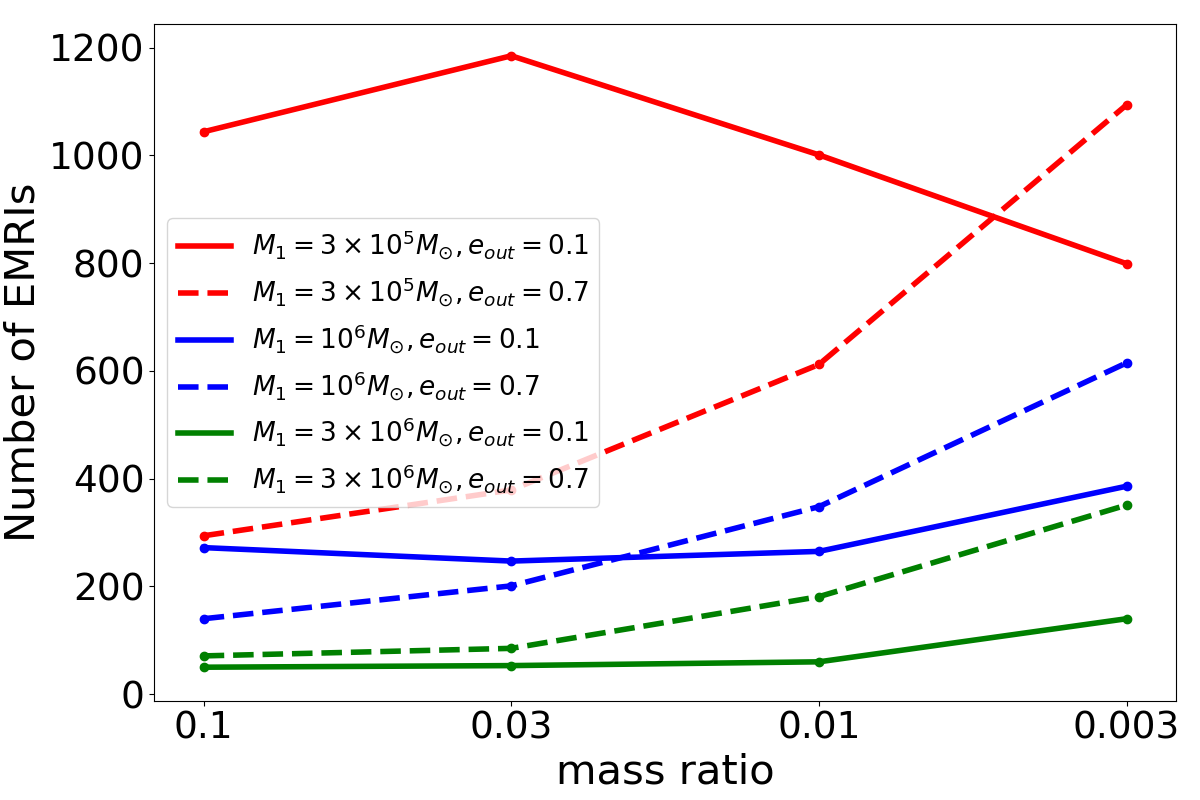}
     \caption{Number of EMRIs as a function of the MBHB's mass-ratio for different masses of the primary and outer eccentricities, as labelled in figure.}
    \label{fig:EMRI_sim}
\end{figure}

In Tab.~\ref{tab:risultati_sim}, we summarise the outcomes of our full sample of three-body simulations. The most likely event is the ejection of the lighter BH, followed by a swap event (in which the stellar-mass BH ends up on a wider orbit, which can be physically thought of as a 'failed ejection'), then direct plunges and finally EMRIs, which occur in 1-5\% of the cases, mainly depending on the primary mass. When considering the MBHB's initial eccentricity, we observe that, for $e_{\rm out}=0.7$, the number of ejections sensibly increases, while the number of DPs decreases. Swap events do not seem to be affected by the initial value of $e_{\rm out}$, while EMRIs seem to form more often in the small-$e_{\rm out}$ case (i.e. when $e_{\rm out}=0.1$), but without a well-defined trend across all of the primary's masses and mass-ratios.
The increasing number of ejections for high $e_{\rm out}$ can be likely explained by the stronger perturbation of the secondary MBH which, with the same $a_{\rm out}$, gets much closer to the inner binary in this case. The strength of the perturbation exerted by $M_2$ also affects the number of EMRIs, as for lower mass-ratios (i.e. weaker perturbing action) we see more EMRIs. This is shown in Fig.~\ref{fig:EMRI_sim}, which depicts the number of EMRIs as a function of the MBHB's mass-ratio for different masses of the primary. Except for $M_1 = 3\times 10^{5} M_{\odot}$ and $e_{ \rm  out} = 0.1$, all other cases show an increasing number of EMRIs when $q$ decreases from 0.1 to 0.003. We also note that for $M_1=3\times 10^{5} M_{\odot}$ the number of EMRIs generated is larger when $e_{\rm out}=0.1$ rather than $e_{ \rm  out} = 0.7$ (except for $q=0.003$), while the opposite is true for $M_1 = 3\times 10^6 M_{\odot}$. The case $M_1 = 10^6 M_{\odot}$ is in between: there is a larger number of EMRIs for $e_{ \rm  out}= 0.1$ when $q \geq 0.03$ while the opposite is true for $q \leq 0.01$.

These observed trends will be discuss and justified in the next sections.

\subsubsection{Timescales}

\begin{figure}
    \centering
    \begin{subfigure}[!h]{1\columnwidth}
        \centering
        \includegraphics[width=1\columnwidth]{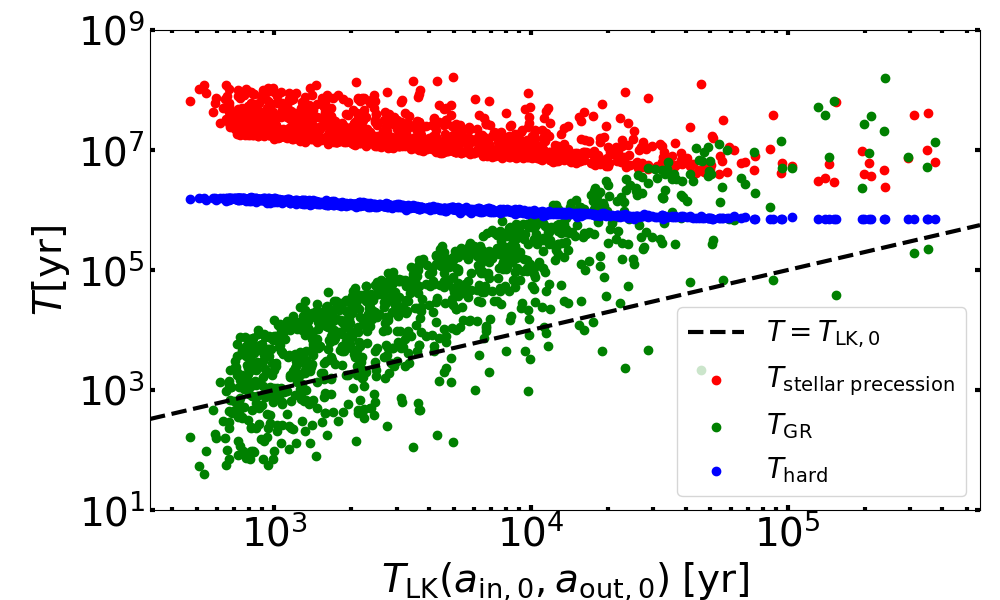}
    \end{subfigure}%
    
    \begin{subfigure}[!h]{1\columnwidth}
        \centering
        \includegraphics[width=1\columnwidth]{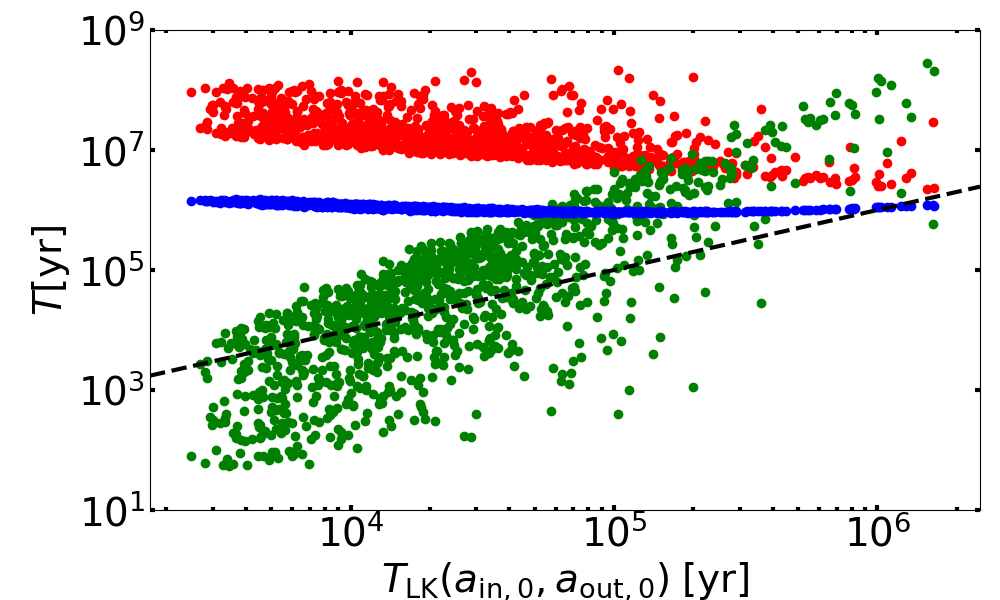}
    \end{subfigure}

    \begin{subfigure}[!h]{1\columnwidth}
        \centering
        \includegraphics[width=1\columnwidth]{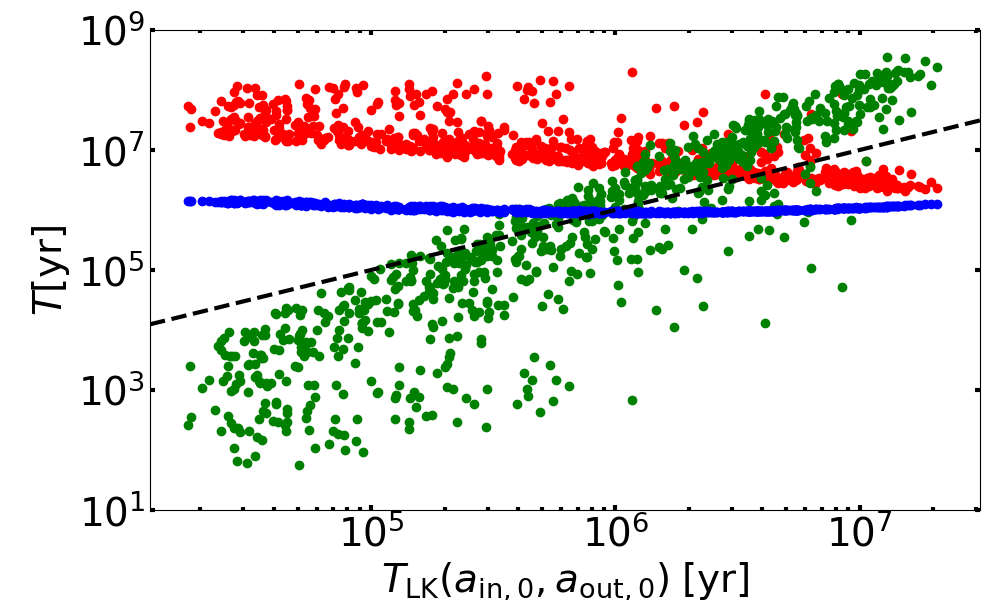}
    \end{subfigure}
    \caption{Comparison between the different timescales of the dynamical mechanisms (stellar hardening, GR, and Newtonian precession) affecting the triplet evolution with respect to the LK one, set as reference on the $x$-axis. All the timescales are computed considering the initial values of the parameters of each individual simulation.
    Red dots represent the Newtonian precession timescale $T_{\rm star}$, green dots the GR precession one $T_{\rm GR}$, while blue symbols denote the stellar hardening time $T_{\rm hard}$. All panels
    feature $M_1 = 3\times 10^5\rm M_{\odot},\,e_{ \rm out}=0.1$, and from top to bottom $q=0.1,\,0.03,\, 0.003$.
    The dashed line marks the equality with the LK timescale $T=T_{ \rm LK}$ where $T$ is any other timescale.
    } 
\label{fig:Timescales}    
\end{figure}
To better understand the main physical processes behind EMRI formation we can investigate the four characteristic timescales associated to the dynamical system under scrutiny. 

The first is the LK timescale, that takes the form \citep[see e.g.][]{10.1093/mnras/stv1552}
\begin{equation}\label{eq:T-q}
    T_{ \rm LK}= \frac{2}{3\pi}\frac{M_1}{M_2}\frac{P_{ \rm out}^2}{P_{ \rm in}}(1-e_{ \rm out}^2)^{3/2}  \propto \frac{1+q}{q} .
\end{equation}
Then, by integrating Eq.~\ref{hard}, we have the hardening timescale, 
\begin{equation}
    T_{ \rm hard} = \frac{\sigma_{ \rm inf}}{G \rho_{ \rm eff,inf}}\int_{ \rm a_{ \rm in,0}}^{a_{ \rm out,0}} \frac{da}{a^2 H(a)},
    \label{eq:hard_tscale}
\end{equation}
that sets the pace at which the secondary's orbit shrinks towards the primary. The shorter the hardening timescale, the shorter the time over which LK oscillations can be effective. Eq.~\eqref{eq:hard_tscale} defines the time needed by the outer binary to harden between its initial semi-major axis $a_{ \rm out,0}$ and $a_{ \rm in,0}$ and is simply obtained by inverting the hardening Eq.~\eqref{hard}.\footnote{Note that the density $\rho$ at the influence radius is evaluated as the artificially increased one used for the simulations. This choice is justified by the fact that we have to compare different timescales considering the effective time in the three-body simulations. The ``density increased'' hardening timescale is, on average, a factor $\approx 10$ shorter than the actual one at a distance of $10^{-2}$ pc from $M_1$. }

Finally, two precession processes affect the dynamics of the system. The first is due to general relativity, acting over a timescale given by:
\begin{equation}
    T_{ \rm GR} = \frac{\pi}{\delta\omega_{ \rm GR}} P_{ \rm in},
\end{equation}
where $\delta\omega_{ \rm GR}$ is the variation of the argument of pericentre of the inner binary due to GR effects, shown in Eq.~\eqref{eq:GR_prec}.
The second is the standard Newtonian precession caused by the non-Keplerian potential of the stellar distribution, acting on a timescale
\begin{equation}
    T_{ \rm star} = \frac{\pi}{\delta\omega_{ \rm star}} P_{ \rm in},
\end{equation}
where $\delta\omega_{\rm star}$ is likewise the variation of the argument of pericentre due to the stellar potential, as from Eq.~\eqref{eq:NW_prec}.

In Fig.~\ref{fig:Timescales} we contrast the LK timescale to  the hardening and precession ones. All timescales are evaluated at the initial conditions of each simulation. From top to bottom we show the trend for decreasing mass-ratio, fixing the outer binary at $M_1 = 3\times 10^5$ and $e_{\rm out} = 0.1$. As expected from Eq.~\eqref{eq:T-q}, the LK timescale becomes longer as $q$ decreases, while the hardening timescale remains approximately constant around $10^6 \rm yr$ for all the cases, being only mildly dependent on $q$. The same holds for the stellar precession timescale.
The hardening timescale is always larger than the LK for $q=0.1$ and $q=0.03$, while for $q=0.003$ there is a consistent fraction of systems where the opposite holds. This means that for those systems the hardening process inhibits secular interactions. In general, instead, the most important process that prevents the LK mechanism from being effective is GR precession. This is likely the reason why the use of an increased effective density, resulting in an artificially shorter hardening timescale, does not appreciably affect the statistics of forming EMRIs (cf. Sec.~\ref{sec:hardening}).

Apart from the $q=0.1$ case, the majority of the EMRIs formed by the lower--mass-ratio MBHB have $T_{ \rm GR}<T_{ \rm LK}$, and thus the EMRI formation via secular interaction should be hindered due to general relativistic effects \citep[see][and references therein]{Naoz2016}. In those cases then we can expect that EMRIs are generally formed via chaotic interactions instead of secular ones, as there is not enough time for the LK oscillations to increase the eccentricity of the stellar BH to the point that an EMRI is formed. GR precession allows the inner binary to remain almost unperturbed until the secondary MBH gets close enough and the triplet enters the chaotic regime.
Another important factor that can inhibit the LK mechanism is the initial relative inclination. Indeed, being the LK effect strongest at $\iota_{\rm rel}\sim 90^{\circ}$ and inhibited for $\iota_{\rm rel}\lesssim 39^{\circ}$, in low--relative-inclination triplets LK oscillations cannot take place. 

Finally, Fig.~\ref{fig:Timescales} also shows that the stellar precession timescale is generally subdominant, which is consistent with the test cases displayed in Appendix~\ref{sec:appA}, where we noticed that the presence or absence of an external stellar potential is not crucial in EMRI formation. As mentioned above the hardening timescale has been artificially shortened in our simulations by increasing the density according to equation \eqref{eq:rhoeff}. Therefore, Newtonian precession might play a role in the case $q=0.003$, where, at large separations, it can be shorter of both the LK and the hardening timescales (cf, lower panel of Fig.~\ref{fig:Timescales}). We will return on this point in section \ref{sec:caveats}.\\
The trends with $q$ of the different timescales shown in Fig.~\ref{fig:Timescales}
are valid also for the other values of $M_1$.

\subsubsection{EMRI features}

\begin{figure}
    \centering
    \includegraphics[width=1\columnwidth]{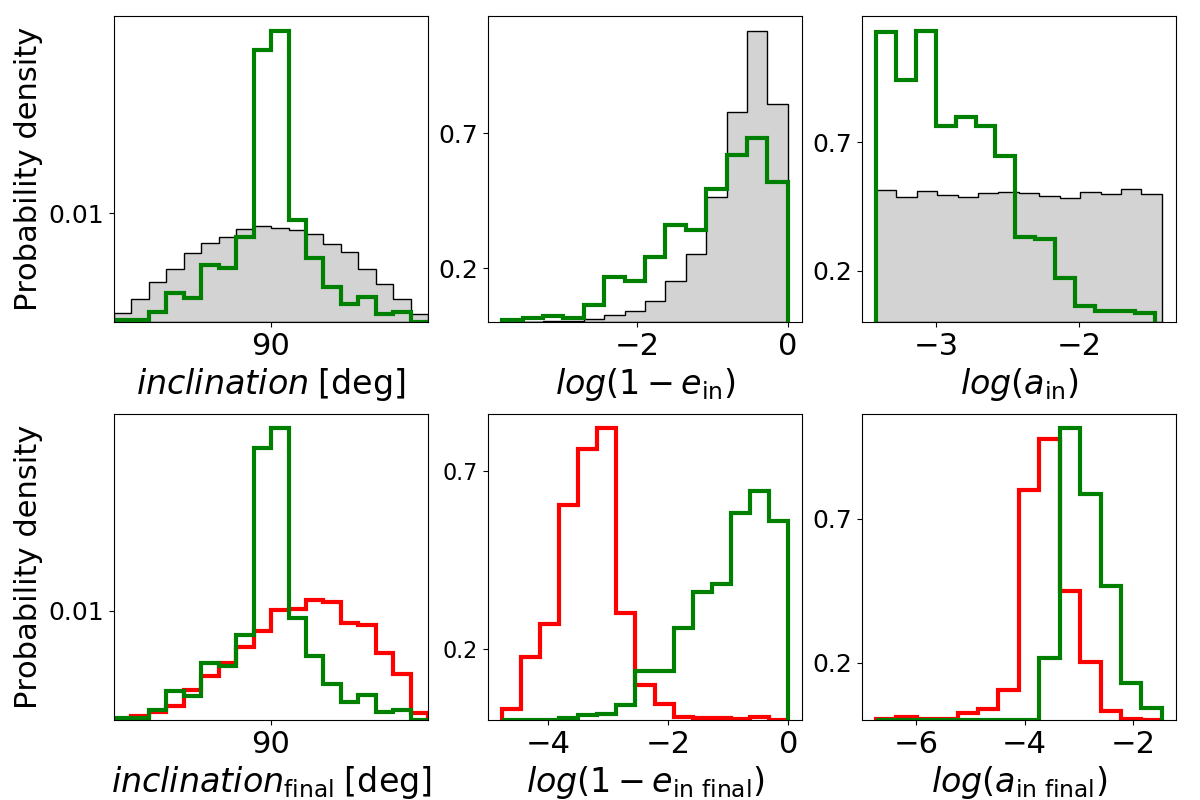}
    \caption{EMRIs relative inclination ($\iota_{\rm rel}$), inner eccentricity ($e_{\rm in}$) and semi-major axis ($a_{\rm in}$) for the case $M_1 = 3\times 10^5 M_{\odot}$, $q= 0.1$, $e_{ \rm out}=0.1$. Upper panels: grey histograms denote the whole sample of initial conditions, while green ones represent the initial parameters of those systems that formed EMRIs. Lower panels: orbital parameters of EMRI systems alone at the beginning (green) and at the end (red) of their three-body simulations.}
    \label{fig:3e5_1e-1_1e-1_E}
    
    \centering
    \includegraphics[width=1\columnwidth]{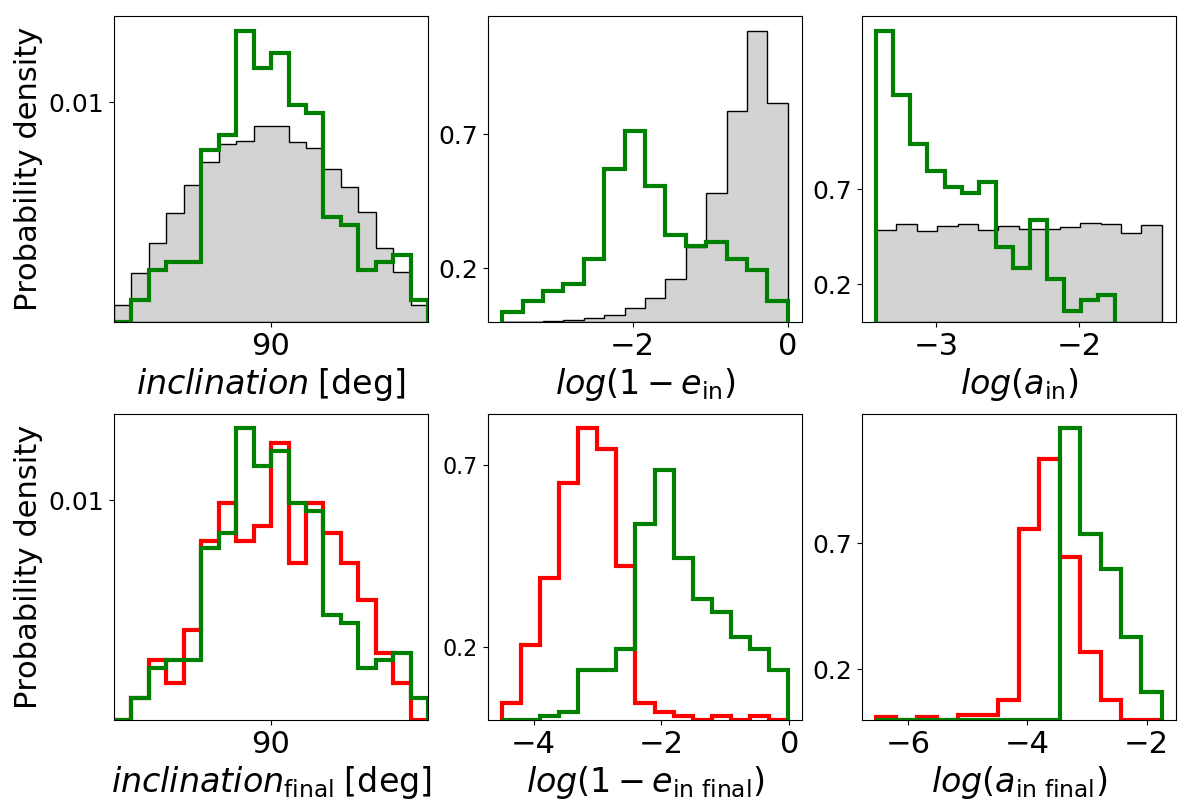}
    \caption{Same as Fig.~\ref{fig:3e5_1e-1_1e-1_E}, but considering the case $M_1 = 3\times 10^5 M_{\odot}$, $q= 0.1$, $e_{ \rm out}=0.7$.}
    \label{fig:3e5_1e-1_7e-1_E}

    \centering
    \includegraphics[width=1\columnwidth]{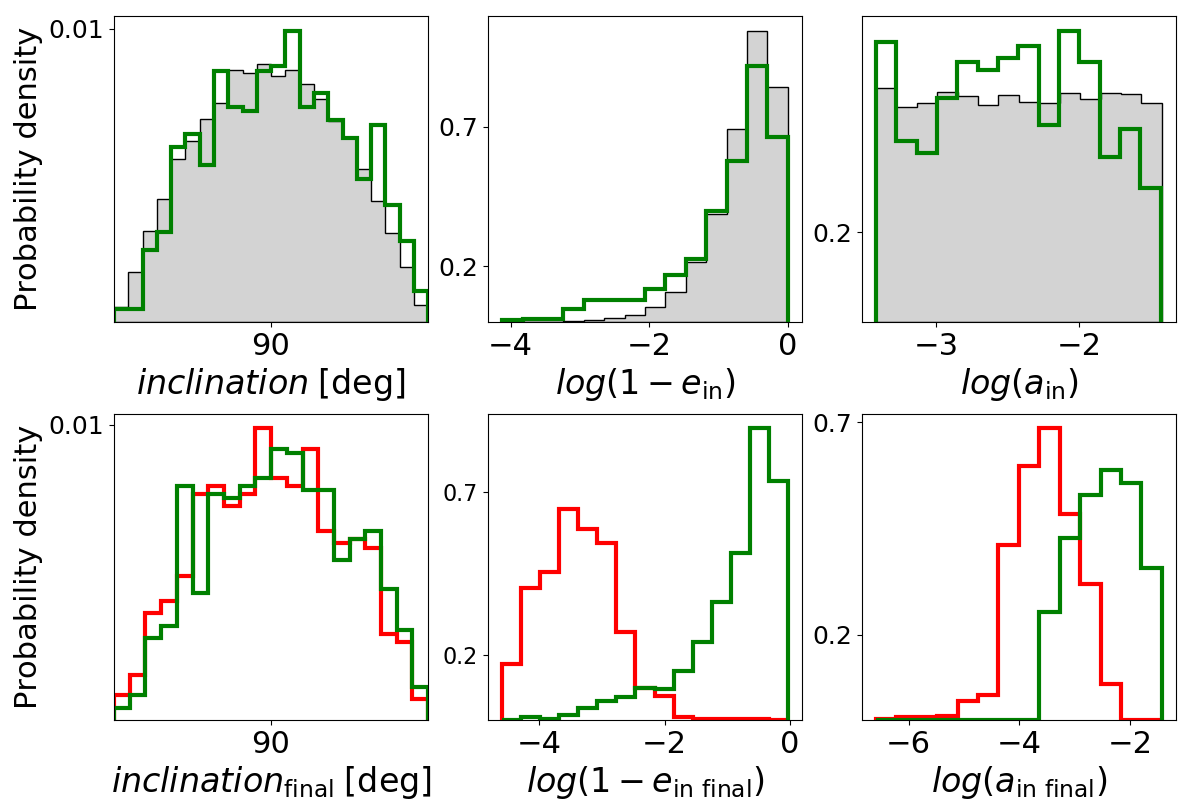}
    \caption{Same as Fig.~\ref{fig:3e5_1e-1_1e-1_E}, but considering the case $M_1 = 3\times 10^5 M_{\odot}$, $q= 0.003$, $e_{ \rm out}=0.1$.}
    \label{fig:3e5_3e-3_1e-1_E}
\end{figure}

In Fig.~\ref{fig:3e5_1e-1_1e-1_E},~\ref{fig:3e5_1e-1_7e-1_E} and~\ref{fig:3e5_3e-3_1e-1_E} we analyse the relation between initial orbital parameters and successful EMRI formation in more details. In each figure, the upper panels show in the grey histograms the initial distributions of \{$\iota,\,1-e_{\rm in},\,\log a_{\rm in}$\} for all 20k simulations (at fixed parameters of the MBHB) compared with the sub-sample of those that yielded an EMRI (green). The lower panels specifically focus on EMRI systems instead, and compare the initial distribution of the same quantities (green) with the final ones (red), where "final" indicates the time where the concerned three-body simulations met the EMRI stoppage condition.

The comparison between the three figures reveals interesting trends.
Starting with the the case $M_1=3\times10^5$, $q= 0.1$, $e_{ \rm out}= 0.1$ (Fig.~\ref{fig:3e5_1e-1_1e-1_E}), we observe that the majority of EMRIs form from systems with high initial relative inclination $\approx 90^{\circ}$, i.e where the LK mechanism is most effective. We therefore expect secular interactions to play a dominant role in EMRI formation (see also Fig.~\ref{fig:Timescales}).
We also note that the eccentricities of the EMRI progenitors do not significantly departure from the initial thermal distribution, whereas there is a strong selection in $a_{\rm in}$. EMRI progenitors tend to be already tightly bound to $M_1$, which makes them less prone to ejection by strong perturbations due to the presence of $M_2$ (that is only 10 times smaller than $M_1$ here).
From the red histograms in the lower panels of Fig.~\ref{fig:3e5_1e-1_1e-1_E}, we
see that these progenitors produce EMRIs preferentially on retrograde orbits. This could be due to a combination of things: as shown by \citet{Miller2002}, when we move out from the test particle limit the threshold for getting a very high eccentricity shifts above $90^\circ$. Moreover when octupole terms are considered (which is the case in our computation, since we are numerically solving the full set of equations), the inner binary can flip from prograde to retrograde orbits and during the flip a very high eccentricity is reached.
These trends are a result of the combined action of LK oscillations and GW emission: during a LK cycle, as the eccentricity grows $\iota$ moves away from $90^\circ$ (either toward 0 or 180 degrees) and when $e_{\rm in}$ reaches its maximum, with values very close to unity, the GW timescales drops abruptly and EMRI formation is detected with the corresponding $\iota$ value. GW emission is also responsible for the decrease in $a_{\rm in}$ as a consequence of energy extraction, which is absent in the standard LK mechanism.

In Fig.~\ref{fig:3e5_1e-1_7e-1_E}, we consider the same configuration of Fig.~\ref{fig:3e5_1e-1_1e-1_E}, except for the higher outer eccentricity, $e_{\rm out} = 0.7$. Looking at the properties of EMRI progenitors (upper panels, green lines), we can still see the peak around $\iota_{\rm rel} \approx 90^\circ$, although less pronounced. The eccentricity distribution instead looks very different from that of Fig.~\ref{fig:3e5_1e-1_1e-1_E}, and specifically it shows a peak around $\log (1-e_{\rm in}) \sim -2$, i.e. now more initially eccentric inner binaries generate EMRIs. The initial semi-major axis distribution shows a bias towards small values similar to what already seen in Fig.~\ref{fig:3e5_1e-1_1e-1_E}. By looking at the lower panel, we see that also in this case at EMRI formation the eccentricity greatly increases and the semi-major axis slightly decreases. EMRI systems are also preferentially retrograde, although much less than the $e_{\rm out} = 0.1$ case. The major difference here is due to the stronger perturbation exerted by $M_2$ which, being on a more eccentric orbit, can get closer to the inner binary. In this situation, chaotic dynamics competes with the LK secular oscillations. Strong chaotic encounters easily eject $m_3$, thus EMRI systems stem from tighter and more eccentric inner binaries, which are less prone to this external perturbation.

Finally, moving to the case $M_1=3\times10^5$, $q= 0.003$, $e_{ \rm out}= 0.1$ (Fig.~\ref{fig:3e5_3e-3_1e-1_E}), we can clearly appreciate how chaotic dynamics almost completely dominates over LK oscillations, as we do not observe any strong selection of high-inclination systems. This is because, at such a small $q$, the LK timescale becomes comparable or longer than the hardening timescale, and as $M_2$ approaches the inner binary, chaotic dynamics becomes efficient in producing EMRIs. Differently from the $q=0.1$ case, since the perturbation is milder, wider inner binaries can end up forming EMRIs too (see the semi-major axis distribution in the upper right panel). Regardless of the main dynamical driver, also in this low-$q$ case EMRIs form when the eccentricity reaches very high values, so that GW emission is more efficient.

\begin{figure}
    \centering
    \begin{subfigure}[!h]{1\columnwidth}
        \centering
        \includegraphics[width=1\columnwidth]{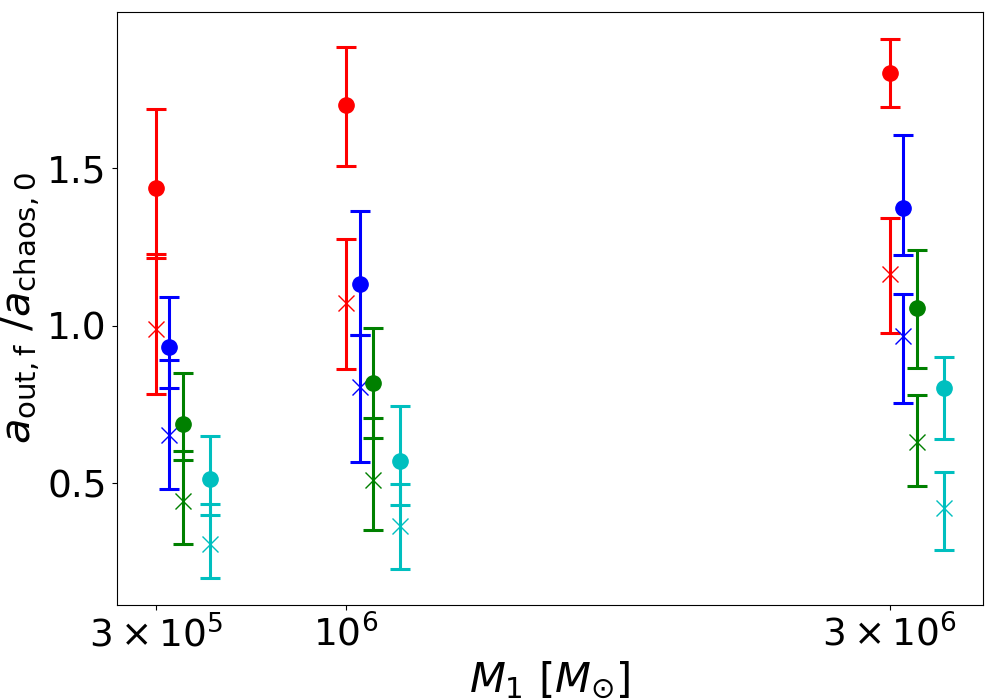}
    \end{subfigure}%
    
    \begin{subfigure}[!h]{1\columnwidth}
        \centering
        \includegraphics[width=1\columnwidth]{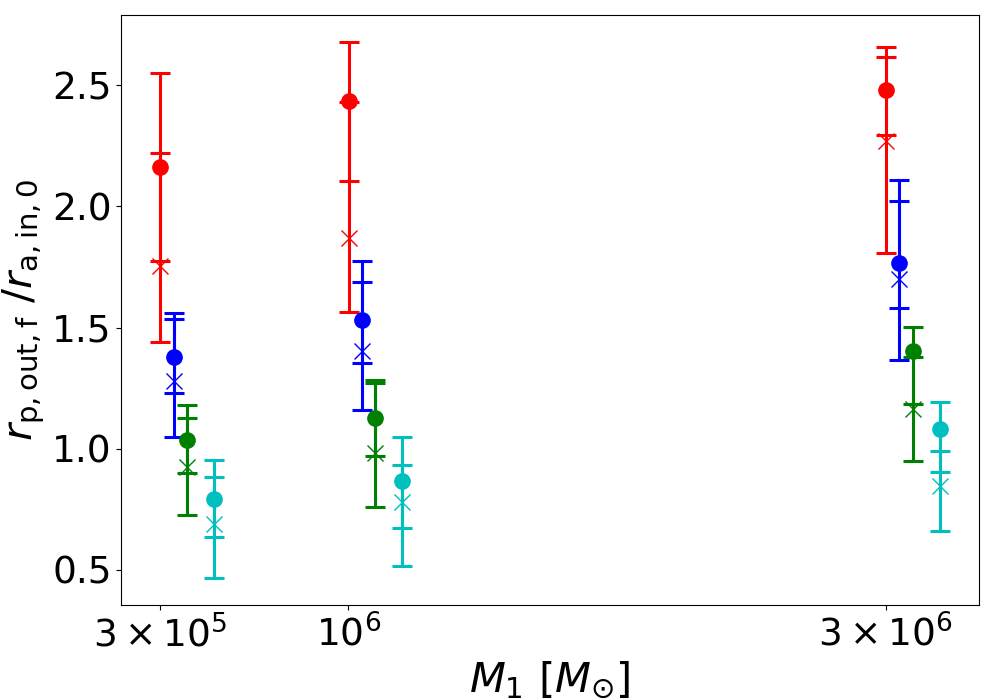}
    \end{subfigure}
    \caption{Upper panel: median values of the ratio $a_{ \rm out,f}/a_{ \rm chaos,0}$ as a function of primary MBH's mass. Different mass-ratios $q$ are shown with different colors: red corresponds to $q=0.1$, blue to $q=0.03$, green to $q=0.01$, cyan to $q=0.003$. Different symbols pertain to eccentricities, a circle for $e_{\rm out}=0.1$ and a cross for $e_{\rm out}=0.7$. The error bars refer to the first and the third quartile of the distributions.
    Lower panel: same as the upper panel, but considering the ratio $r_{ \rm p,out,f}/r_{ \rm a,in,0}$. In both panels, points pertaining to different mass-ratios have been arbitrarily shifted along the $x$-axis to ease readability.}
\label{fig:chaos}    
\end{figure}

An additional indicator of the prevalence of either the secular or the chaotic nature of the EMRI formation process is given by the ratio $a_{ \rm out,f}/a_{ \rm chaos,0}$, shown in Fig.~\ref{fig:chaos} as a function of $M_1$. $a_{ \rm chaos,0}$ represents the minimum value of $a_{ \rm out}$ for which the initial configuration of the three bodies is stable, while $a_{ \rm out,f}$ is the value of $a_{ \rm out}$ at the end of the simulation. 
If LK secular oscillations operate on a sufficiently short timescale, then an EMRI can form before $a_{ \rm out}$ reaches the instability threshold, resulting in secular-driven EMRIs characterised by $a_{ \rm out,f}/a_{ \rm chaos,0} \geq 1$. On the contrary, if the secular interaction is not strong enough, the hardening process will carry $M_2$ down to small values of $a_{ \rm out}$ until chaotic interactions take over the dynamics.
From the same figure, the ratio is generally higher for $e_{\rm out}=0.1$ with respect to $e_{\rm out}=0.7$, and a clear decreasing trend of the ratio for decreasing values of $q$ is observable. There is also a slight increase of the value of $a_{ \rm out,f}/a_{ \rm chaos,0}$ with increasing $M_1$, showing that chaotic interactions are favoured for lighter primary masses.

In the lower panel of Fig.~\ref{fig:chaos}, we show the ratio $r_{ \rm p,out,f}/r_{ \rm a,in,0}$, i.e. the ratio between the final pericentre of the outer binary and the initial apocentre of the inner one. Again if this ratio is less than 1, we have an indication that the interactions are mainly chaotic, while if greater, it is more likely that secular effects led to the EMRI formation. Indeed, if the final pericentre of the outer binary is larger than the initial apocentre of the inner one, it means that the secondary MBH never crossed the orbit of the stellar BH, and so the chaotic interactions are likely subdominant. Also here, more circular and higher-$q$ MBHBs tend to produce EMRIs through secular interactions, while lower-$q$ and more eccentric ones determine a prevalence of chaotic interactions.

\subsection{EMRI Formation Burst} \label{sec: sis rate 2}
\begin{table}
    \setlength\extrarowheight{3pt}
    \centering
    \begin{tabular}{cc|cc|c}
        \toprule
        \multicolumn{2}{c|}{$M_1 = 3\times 10^5 M_{\odot}$} & \multirow{2}{*}{$\mathcal{N}_{\text{EMRI,SIS}}$} & \multirow{2}{*}{$\left\langle T_{\text{tot}}\right\rangle\;\rm \left[yr\right]$} & \multirow{2}{*}{Rates $\left[\rm yr^{-1}\right]$} \\
        $q$ & \multicolumn{1}{c|}{$e_{\text{out},0}$} &  &  &  \\
        \midrule
        \multirow{2}{*}{$0.1$} & \multicolumn{1}{c|}{$0.1$} & $43.7$ & $5.5\times10^6$ & $2\times10^{-5}$ \\
                               & \multicolumn{1}{c|}{$0.7$} & $11.4$ & $3.7\times10^6$ & $5\times10^{-6}$ \\
        \hline
        \multirow{2}{*}{$0.03$} & \multicolumn{1}{c|}{$0.1$} & $60.7$ & $1.1\times10^7$ & $10^{-5}$ \\
                                & \multicolumn{1}{c|}{$0.7$} & $23.3$ & $7.7\times10^7$ & $5\times10^{-6}$ \\
        \hline
        \multirow{2}{*}{$0.01$} & \multicolumn{1}{c|}{$0.1$} & $82.3$ & $1.6\times10^7$ & $8\times10^{-6}$ \\
                                & \multicolumn{1}{c|}{$0.7$} & $45.0$ & $1.7\times10^7$ & $4\times10^{-6}$ \\
        \hline
        \multirow{2}{*}{$0.003$} & \multicolumn{1}{c|}{$0.1$} & $100.2$ & $3.9\times10^7$ & $4\times10^{-6}$ \\
                                 & \multicolumn{1}{c|}{$0.7$} & $93.2$ & $4.4\times10^7$ & $2\times10^{-6}$ \\
        \bottomrule
    \end{tabular}
    \begin{tabular}{cc|cc|c}
        \toprule
        \multicolumn{2}{c|}{$M_1=10^6 M_{\odot} \ \ \ \ \ \ \ $} & \multirow{2}{*}{$\mathcal{N}_{\text{EMRI,SIS}}$} & \multirow{2}{*}{$\left\langle T_{\text{tot}}\right\rangle\;\rm \left[yr\right]$} & \multirow{2}{*}{Rates $\left[\rm yr^{-1}\right]$} \\
        $q$ & \multicolumn{1}{c|}{$e_{\text{out},0}$} &  &  &  \\
        \midrule
        \multirow{2}{*}{$0.1$} & \multicolumn{1}{c|}{$0.1$} & $45.6$ & $4.3\times10^6$ & $2\times10^{-5}$ \\
                               & \multicolumn{1}{c|}{$0.7$} & $22.6$ & $5.2\times10^6$ & $6\times10^{-6}$ \\
        \hline
        \multirow{2}{*}{$0.03$} & \multicolumn{1}{c|}{$0.1$} & $58.9$ & $8.8\times10^6$ & $10^{-5}$ \\
                                & \multicolumn{1}{c|}{$0.7$} & $41.5$ & $1.0\times10^7$ & $6\times10^{-6}$ \\
        \hline
        \multirow{2}{*}{$0.01$} & \multicolumn{1}{c|}{$0.1$} & $127.9$ & $1.8\times10^7$ & $10^{-5}$ \\
                                & \multicolumn{1}{c|}{$0.7$} & $88.8$ & $2.5\times10^7$ & $4\times10^{-6}$ \\
        \hline
        \multirow{2}{*}{$0.003$} & \multicolumn{1}{c|}{$0.1$} & $222.5$ & $5.7\times10^7$ & $5\times10^{-6}$ \\
                                 & \multicolumn{1}{c|}{$0.7$} & $218.2$ & $6.6\times10^7$ & $4\times 10^{-6}$ \\
        \bottomrule
    \end{tabular}
    \begin{tabular}{cc|cc|c}
        \toprule
        \multicolumn{2}{c|}{$M_1 = 3\times 10^6 M_{\odot}$} & \multirow{2}{*}{$\mathcal{N}_{\text{EMRI,SIS}}$} & \multirow{2}{*}{$\left\langle T_{\text{tot}}\right\rangle\;\rm \left[yr\right]$} & \multirow{2}{*}{Rates $\left[\rm yr^{-1}\right]$} \\
        $q$ & \multicolumn{1}{c|}{$e_{\text{out},0}$} &  &  &  \\
        \midrule
        \multirow{2}{*}{$0.1$} & \multicolumn{1}{c|}{$0.1$} & $25.1$ & $4.3\times10^6$ & $9\times10^{-6}$ \\
                               & \multicolumn{1}{c|}{$0.7$} & $34.9$ & $6.4\times10^6$ & $8\times10^{-6}$ \\
        \hline
        \multirow{2}{*}{$0.03$} & \multicolumn{1}{c|}{$0.1$} & $59.9$ & $8.9\times10^6$ & $10^{-5}$ \\
                                & \multicolumn{1}{c|}{$0.7$} & $54.0$ & $1.2\times10^7$ & $7\times10^{-6}$ \\
        \hline
        \multirow{2}{*}{$0.01$} & \multicolumn{1}{c|}{$0.1$} & $97.0$ & $2.1\times10^7$ & $7\times10^{-6}$ \\
                                & \multicolumn{1}{c|}{$0.7$} & $218.4$ & $3.1\times10^7$ & $10^{-5}$ \\
        \hline
        \multirow{2}{*}{$0.003$} & \multicolumn{1}{c|}{$0.1$} & $274.9$ & $7.2\times10^7$ & $6\times10^{-6}$ \\
                                 & \multicolumn{1}{c|}{$0.7$} & $362.1$ & $9.8\times10^7$ & $5\times10^{-6}$ \\
        \bottomrule
    \end{tabular}
    \caption{EMRI rates for different mass of the primary, mass-ratio and outer eccentricity of the MBHB, assuming a SIS-like initial BH cusp around $M_1$.}
    \label{tab:EMRI_rates}
\end{table}

Following the procedure outlined in Section~\ref{sec:sis rate 1}, we report in Tab.~\ref{tab:EMRI_rates} the EMRI rate inferred from our simulation pool of 24 different combinations of $M_1$, $q$, $e_{\rm out}$. The rates are evaluated by dividing the number of formed EMRIs (assuming a SIS-like density cusp around $M_1$) by the average total time taken by the outer binary to shrink down to $a_{\rm out,f}$ plus the time between the EMRI identification and its plunge into $M_1$ (cf. Eq.~\eqref{T_tot}). 
From the table we note that the number of EMRIs $\mathcal{N}_{ \rm EMRIs,SIS}$, for every value of $M_1$, grows with decreasing mass-ratio $q$, reaching its maximum value at $q=0.003$. For the cases $M_1=3\times10^5 \rm M_{\odot}$ and $10^6$ $M_{ \rm \odot}$, $\mathcal{N}_{ \rm EMRIs,SIS}$ is always higher when considering an almost circular MBHB with respect to an eccentric one, while the opposite is true for the case $M_1= 3 \times 10^6 \rm M_{\odot}$, except for $q=0.01$ where, however, the difference is very small.  
In particular looking at circular MBHBs, the number of EMRIs in a SIS increases going from $q=0.1$ to $q=0.003$. Specifically, the increase is a factor $\sim 2$ for $M_1=3\times10^5 \rm M_{\odot}$, a factor of $\sim 5$ for $M_1=10^6 \rm M_{\odot}$ and a factor of $\sim 10$ for $M_1=3\times10^6 \rm M_{\odot}$. For eccentric MBHBs, instead, the increasing factor is always $\sim 10$ when shifting from high to low mass-ratios of the MBHB, irrespective of $M_1$.

Concerning the EMRI formation rate, we get that for all the cases considered it is between $10^{-5}$ and $10^{-6} \rm \ yr^{-1}$, i.e. a factor $10-100$ larger than the EMRI rates reported in the literature for the standard two-body formation channel. We can therefore conclude that MBHBs are efficient triggers of EMRI events. The rate appears to weakly depend on the mass of the primary $M_1$ and on the eccentricity of the MBHB, decreasing slightly with increasing $M_1$ and for higher $e_{ \rm out}$.

\begin{figure}
    \centering
    \includegraphics[width=1\columnwidth]{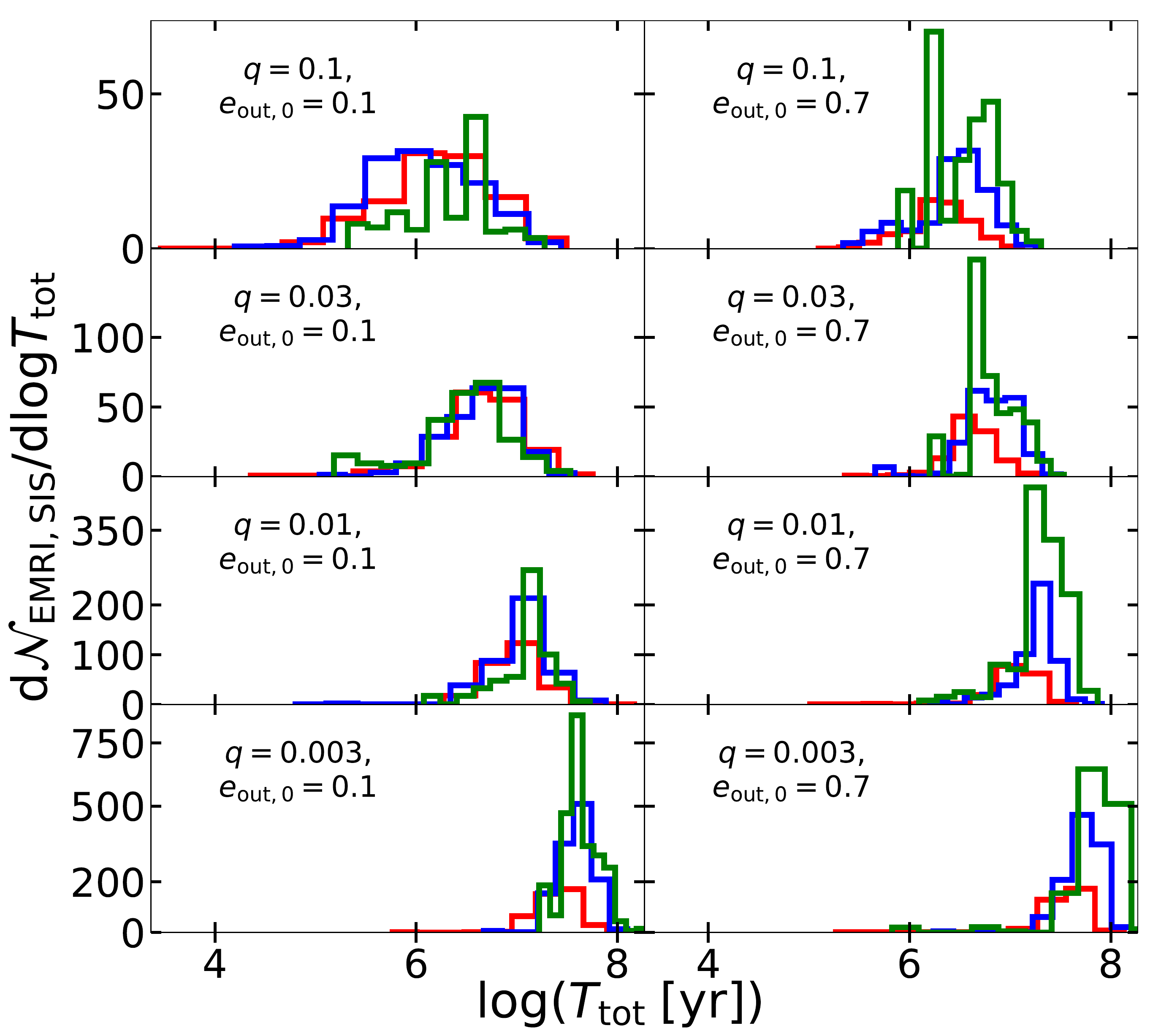}
    \caption{Time distribution of EMRI formation once re-scaled to a SIS cusp. The three colors refer to the cases: red for $M_1=3\times10^5\rm M_{ \rm \odot}$, blue for $M_1=10^6\rm M_{ \rm \odot}$, green for $M_1=3\times10^6\rm M_{ \rm \odot}$.}
\label{fig:time_rate}    
\end{figure}

Although the inferred rates are large, we should keep in mind that these events occur over a timescale comparable to the MBHB's hardening time, typically shorter than the Hubble time by a factor $\lsim 100$. This is clearly shown in Fig.~\ref{fig:time_rate}, which shows the differential number of EMRIs as a function of time, ${d\mathcal{N}_{ \rm EMRIs,SIS}}/({d\log T_{\rm tot}})$, produced in the process.

We see that the rate peaks at different times depending on the properties of the MBHB. 
The time needed by an EMRI to form via the MBHB-driven channel is always between $10^{5}$ and $10^{8}$ yr. In particular the formation time becomes longer as the mass-ratio $q$ becomes smaller, and slightly grows with the mass of the primary MBH. Moreover the time lapse in which EMRIs are formed does not strongly depend on the MBHB's eccentricity for fixed values of $q$ and $M_{1}$. 

We can now summarize some of the result obtained. Considering the case $e_{\rm out}=0.1$ representative, we can say that low-eccentricity MBHBs with high $q$ promote fast EMRI formation mainly driven by the LK secular mechanism. The total number of EMRIs formed tends, however, to be small. Conversely, since for low $q$ LK is less efficient and the evolution is driven by the hardening timescale, EMRI formation takes longer, but produces a larger number of events.
Moreover, for low values of $q$, GR precession also inhibits LK oscillations, and EMRIs are generally formed after the hardening process has taken $M_2$ down to low values of $a_{\rm out}$, thus generating a chaotic triplet. For $e_{\rm out}=0.7$, instead, EMRI formation at all values of $q$ proceeds mostly via chaotic interactions arising when the secondary MBH passes at its pericentre. Also, in this case the number of EMRIs decreases with increasing $q$, because heavier secondaries exert a stronger pull on the stellar BH facilitating ejection rather than capture.

The fact that, for all the cases considered, there is a burst of EMRI formation on a timescale between $10^{6}$ yr and $10^{8}$ yr is physically motivated by the evolution of the MBHB: we can expect an EMRI burst in coincidence with the typical time spent by a MBHB to shrink from a large separation. However, as the outer semi-major axis decreases, stellar-mass BHs are preferentially ejected due to the system being non-hierarchical, which leads to cusp erosion. Eventually, when the MBHB merges, the process ceases: without a secondary MBH, stellar-mass BHs cannot be directed toward the primary BH more efficiently than through standard 2-body relaxation anymore. 

The different trends in the EMRI numbers with $M_1$ and $q$ appearing in Tab.~\ref{tab:EMRI_rates} compared to Tab.~\ref{tab:risultati_sim} stem from the procedure of adapting our simulation to a SIS profile scaling with the central MBH mass (cf. Sec.~\ref{sec:sis rate 1}).
The SIS re-scaling naturally gives more weight to EMRIs formed from BHs initially at larger $a$.
This explains the larger amount of re-scaled EMRIs for the case $M_1= 3\times10^6 M_{ \rm \odot}$ and at lower values of $q$. The mass dependence originates from the fact that the initialization of $a_{\rm in}$ scales with $\beta = (M_{1}/10^{6}M_{\rm \odot})$ and so for heavier $M_1$ the inner binary's semi-major axis range relevant to the EMRI formation process is larger.
Moreover, by comparing Fig.~\ref{fig:3e5_3e-3_1e-1_E} and Fig.~\ref{fig:3e5_1e-1_1e-1_E}, it is evident that for $q=0.003$ EMRI progenitors are distributed all over the initial total distribution of $a_{\rm in}$, while for $q=0.1$ there is a clear selection for lower values of $a_{\rm in}$, thus carrying less weight in the re-scaling procedure.

\subsection{Cosmological Rate}\label{sec: cosmological rate 2}

\begin{figure}
    \centering
    \begin{subfigure}{1\columnwidth}
        \centering
        \includegraphics[width=1\columnwidth]{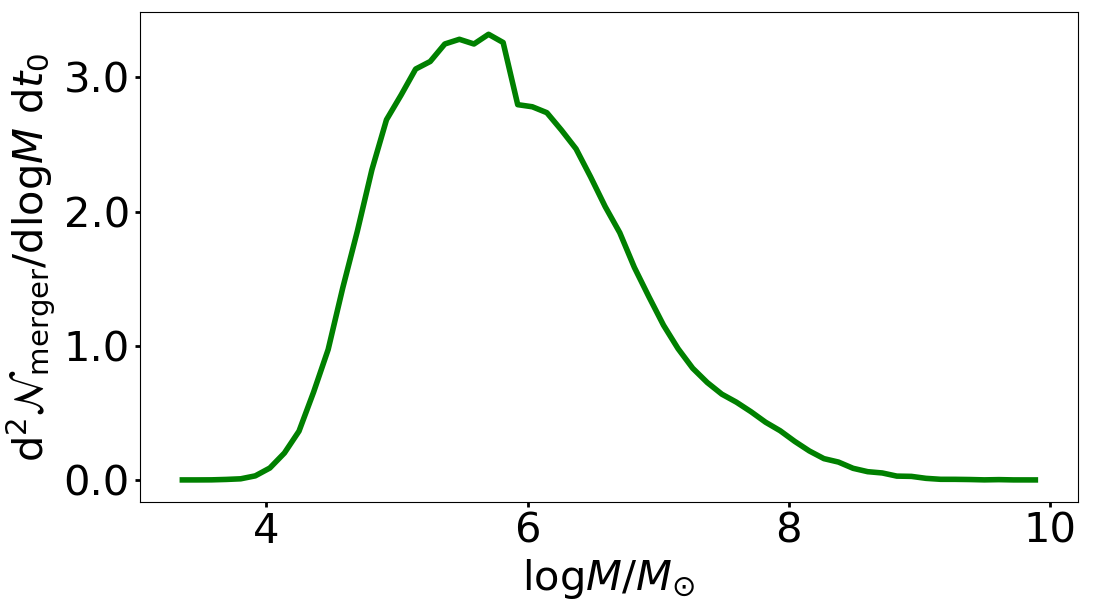}
    \end{subfigure}%
    
    \begin{subfigure}{1\columnwidth}
        \centering
        \includegraphics[width=1\textwidth]{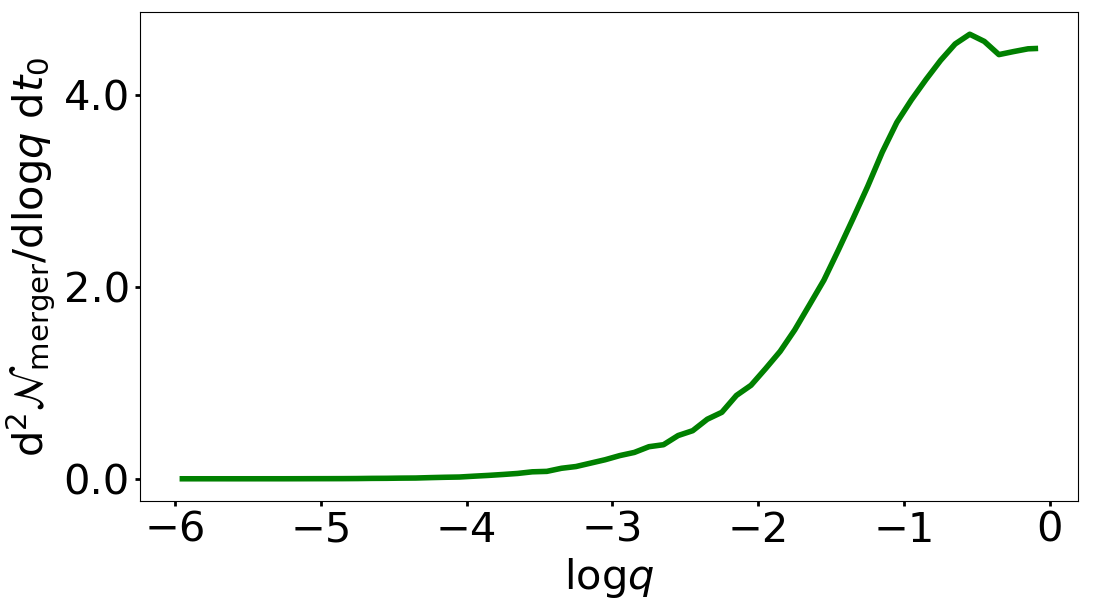}
    \end{subfigure}

    \begin{subfigure}{1\columnwidth}
        \centering
        \includegraphics[width=1\textwidth]{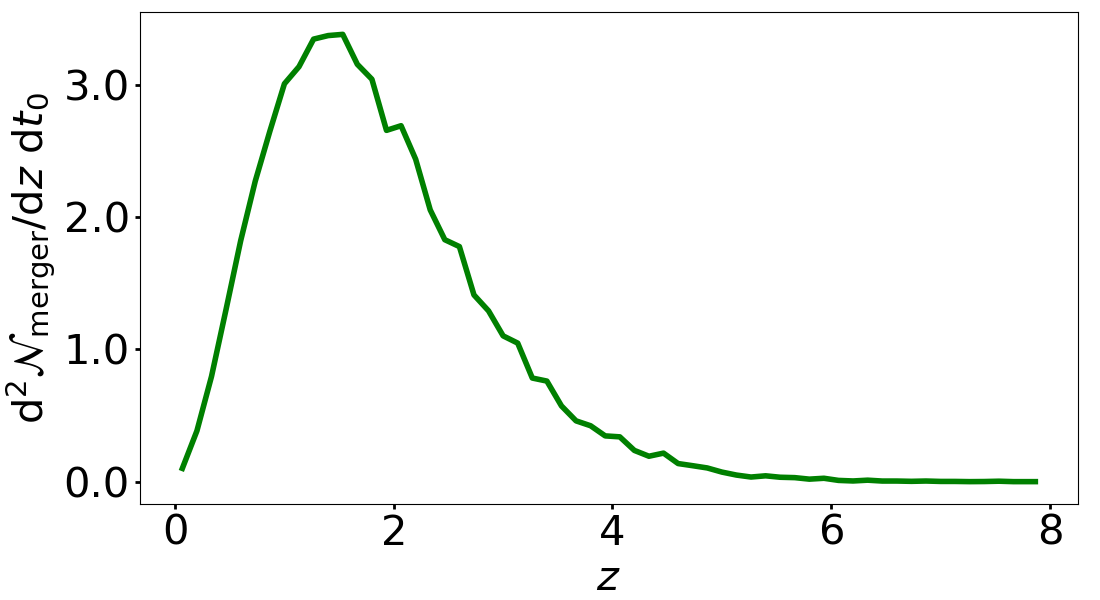}
    \end{subfigure}
    \caption{MBHB merger rate distribution as a function of the primary mass (top), the mass-ratio (middle), and redshift (bottom). 
    } 
\label{fig:rate_distr}    
\end{figure}

As described in Section~\ref{sec:computation of emri rate}, in order to compute the cosmological merger rate of EMRIs we need to couple the results of our three-body simulations, which tell us the efficiency of EMRI formation for a single MBHB, together with a model for the distribution of MBHBs in the Universe. This latter information is extracted from the output of the SAM \textit{L-Galaxies}, whose MBHB merger rate as a function of mass of the primary, mass-ratio and redshift is shown in Fig.~\ref{fig:rate_distr} (from top to bottom). The mass function of the merging binaries has a clear peak at $M_1 = 10^{6} M_{\rm \odot}$ and drops quickly below $M_1 = 10^{5}M_{\rm \odot}$, meaning that the major contributors to the observable rate are MBHBs with a mass of the primary between $10^5 - 10^6 M_{\rm \odot}$. The distribution of the rate over the mass-ratio presents a plateau in the range $q\in[0.1,1]$ and it drastically drops off below $q=0.01$.  This might be partially due to the resolution limit of the simulation, although a similar trend is also seen in Press\&Schechter merger tree-based semi-analytic model that do not suffer such resolution limitations \citep[see e.g.][]{2007MNRAS.377.1711S,2012MNRAS.423.2533B}. Finally the redshift distribution peaks around $z=2$, when the Universe was approximately $\sim 3$ Gyr old.

The global MBHB merger rate as observed from Earth is given integrating Eq.~\eqref{eq:m_rate_2}, i.e.:
\begin{equation}
    \frac{dN}{dt_{\rm 0}}= \int \frac{d^4 N}{d\log M \ d\log q \ dz \ dt_{\rm 0}} d\log M_{\rm 1} d\log q dz.
\end{equation}
For the MBHB population that we selected the result is $\approx 7$ mergers per year.

Then, coupling the number of EMRIs in a SIS cusp obtained in Section~\ref{sec: sis rate 2} with the MBHB merger rate given by Eq.~\eqref{eq:m_rate_2}, we computed the number of EMRIs per unit of $M_1$, $q$, $z$ and time, triggered by MBHBs along the cosmic history, namely Eq.~\eqref{eq:rate_tot}.
Marginalising Eq.~\eqref{eq:rate_tot} over two of the three variables, we obtain the EMRI merger rate as a function of $M_1$, $q$ and $z$ separately.
We distinguished between distributions originated by MBHBs with eccentricity $e_{\rm out}=0.7$ and $e_{\rm out}=0.1$ by assuming that all binaries in the merger tree have either a small (0.1) or large eccentricity (0.7).

\begin{figure}
    \centering
    \begin{subfigure}{1\columnwidth}
        \centering
        \includegraphics[width=1\columnwidth]{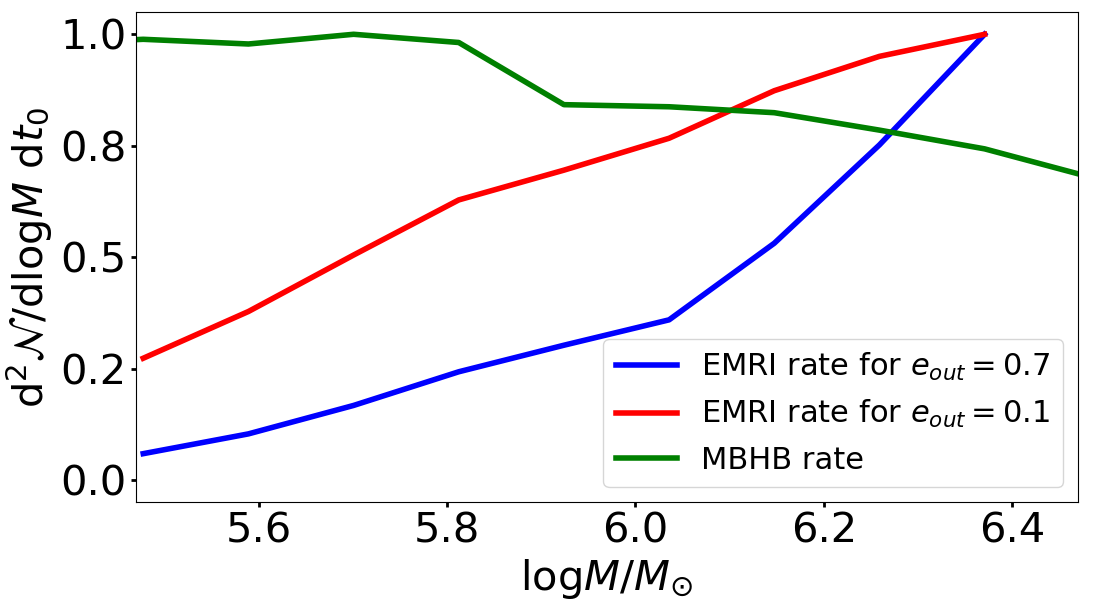}
    \end{subfigure}%
    
    \begin{subfigure}{1\columnwidth}
        \centering
        \includegraphics[width=1\columnwidth]{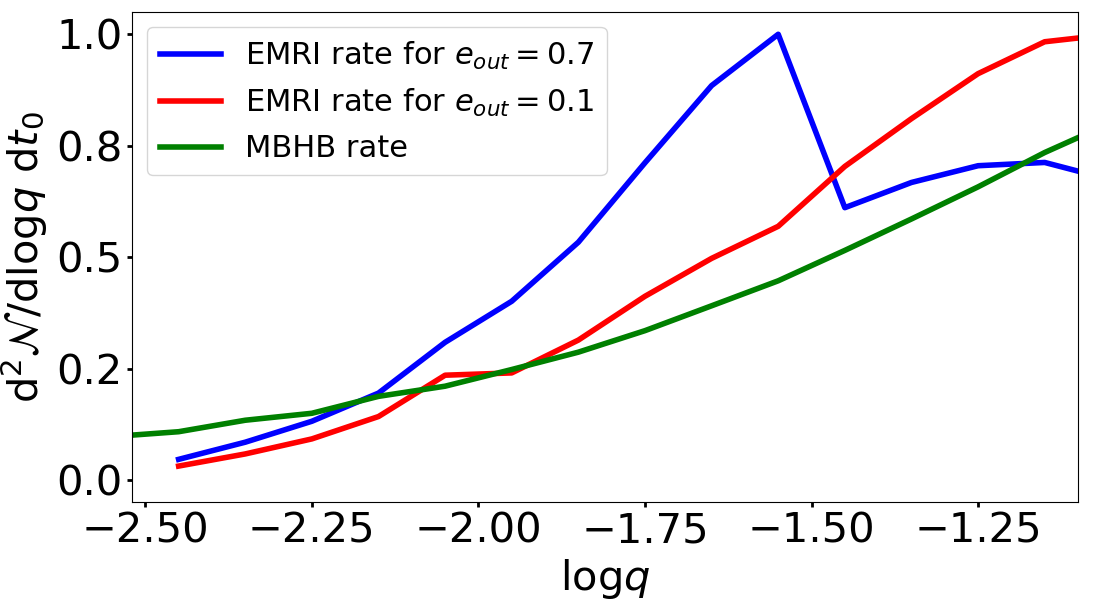}
    \end{subfigure}

    \begin{subfigure}{1\columnwidth}
        \centering
        \includegraphics[width=1\columnwidth]{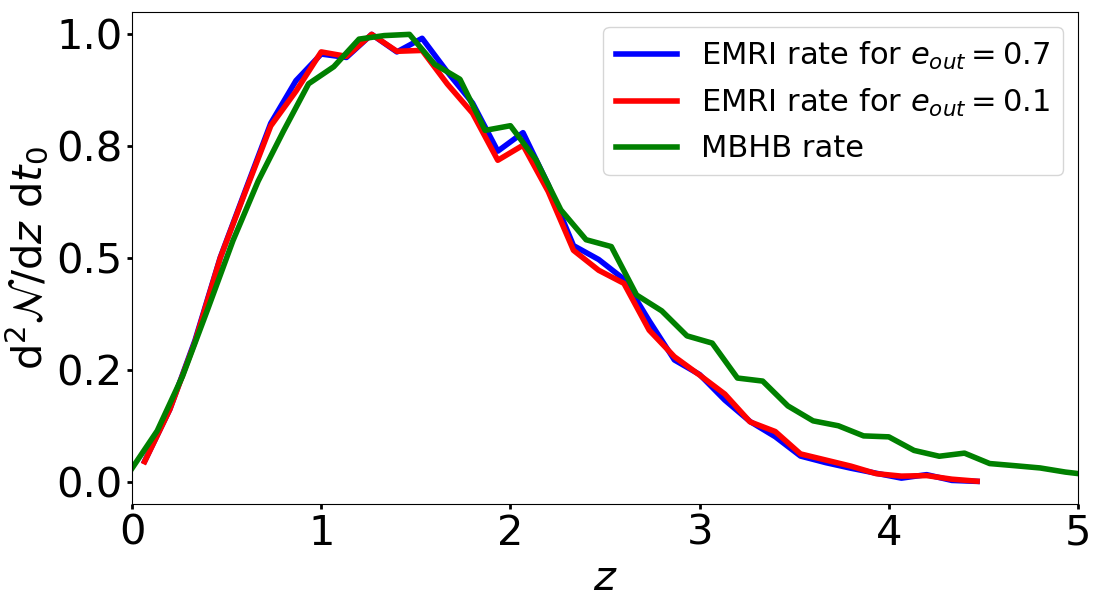}
    \end{subfigure}
    \caption{EMRI rate distribution as a function of the primary mass (top), the mass-ratio (middle), and redshift (bottom) considering the two different initial outer eccentricities: red for $e_{\rm out}=0.1$, blue for $e_{\rm out}=0.7$. Green lines denote the same distribution, but for the MBHB merger rate. All distributions are normalized in order to compare the trend with the MBHB coalescence rate.
    }
\label{fig:fr}    
\end{figure}

Results are shown in Fig.~\ref{fig:fr}.
Looking at the cosmological EMRI formation distribution over redshift, we note that for both the  $e_{\rm out}=0.7$ and $e_{\rm out}=0.1$ cases, the peak is between $z=1$ and $z=2$. This is a direct consequence of the trend of the MBHB merger rate distribution over $z$, since the EMRI rate given by our simulations does not depend on the redshift. 
Considering the variation with the mass-ratio $q$, for almost circular MBHBs the rate is a monotonically increasing function of the decreasing mass-ratio, while for eccentric MBHBs the global rate has a peak around $q=0.03$ and then it decreases and flattens.
Considering the primary MBH $M_1$, the rate shows for both eccentricities an increasing trend for increasing $M_1$, meaning that the major contribution to the total rate comes from those MBHBs with a heavier primary MBH.
We note that the range of $M_1$ and $q$ we simulated does not fully cover the interesting parameter space of EMRI formation. In particular, a large fraction of EMRIs is still formed for $M_1 > 3\times 10^6 M_{\rm \odot}$ and at $q>0.1$. The $M_1$ range was calibrated to produce EMRIs in the LISA sensitivity band sweet spot, and we expect heavy-primary EMRIs to have a small contribution to the LISA observed rate, as shown below. Conversely, the $q=0.1$ cutoff was informed by \cite{2011ApJ...729...13C} findings about TDE events. In that paper, the authors find that TDEs in MBHBs are already significantly suppressed at $q=0.1$. Although also the EMRI rate decreases with increasing $q$, MBHBs with $q>0.1$ can significantly contribute to the overall observed LISA rate.
We do not attempt here to extrapolate rates outside the simulated $M_1$ and $q$ range, but we will discuss below how extending this domain might influence our conclusions.

Lastly, because of the above mentioned parameter space cuts, the values of the rates that we obtained can be considered as conservative.
Integrating Eq.~\eqref{eq:rate_tot} over the three variables $M_1$, $q$, $z$, and distinguishing between $e_{\rm out}=0.1$ and $e_{\rm out}=0.7$, we obtain that the global EMRI merger rate is $\sim 74 \rm yr^{-1}$ and $\sim 64 \rm yr^{-1}$, respectively.

\subsection{Expected LISA detection}\label{sec:Lisa rate 2}

As a final step we determine how many EMRIs triggered by the presence of a MBHB could be detected by LISA. To this end, we performed a Monte Carlo sampling of the distribution given by Eq.~\eqref{eq:rate_tot} assuming a 4 yr LISA mission, specifically obtaining 296 and 256 EMRIs for $e_{\rm out}=0.1$ and $e_{\rm out}=0.7$, respectively. We repeated this procedure 10 times, averaging the results to smooth out potential features due to low-number statistics.
For each event we compute the SNR employing the waveform model implemented in \citet{Bonetti2020}.

\begin{figure}
    \centering
    \includegraphics[width=1\linewidth]{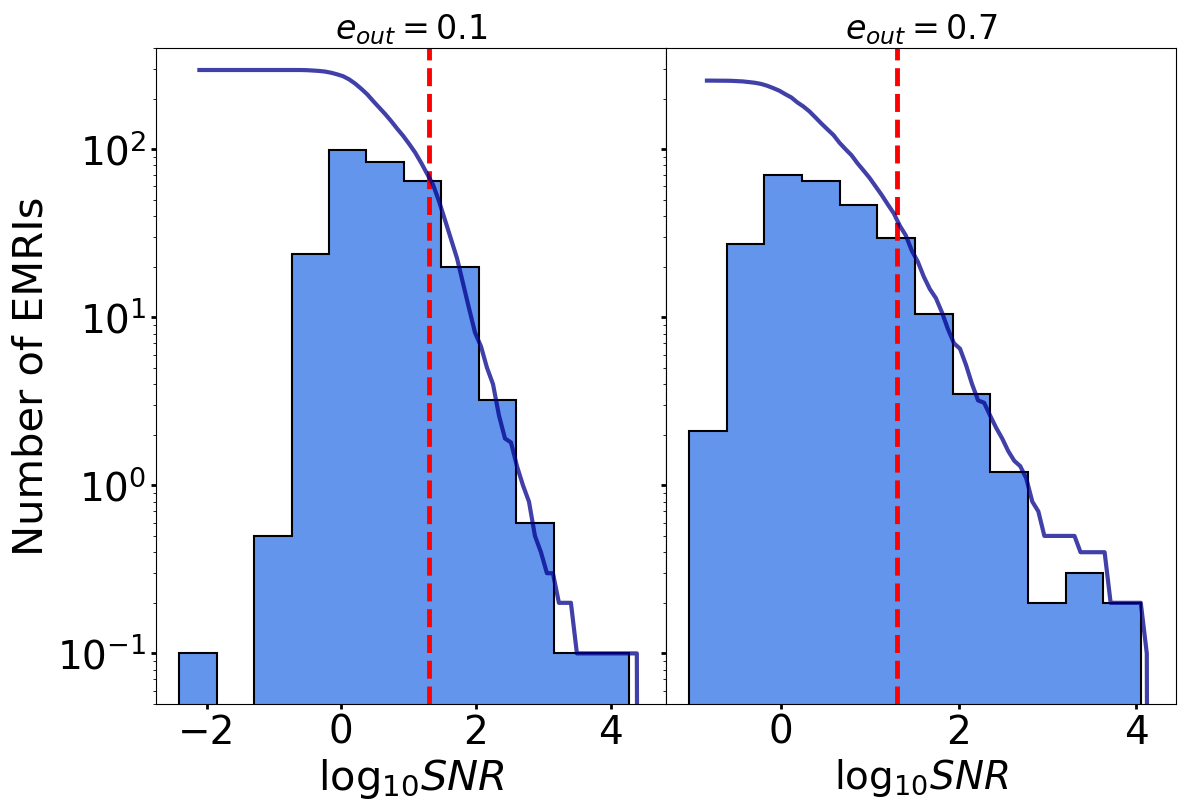}
    \caption{SNR distribution of the number of EMRIs originated in a MBHB with eccentricity $e_{\rm out}=0.1$ (left) and $e_{\rm out}=0.7$ (right) during 4 years of LISA operation. The vertical dashed lines denote the SNR detection threshold of 20, while the dark blu one represents the cumulative distribution of the EMRIs.
    }
    \label{fig:SNR}
\end{figure}

In Fig.~\ref{fig:SNR}, we show the histogram representing the number of EMRIs triggered by a MBHB that are expected to occur in the 4 years of the LISA mission as a function of the SNR of their GW emission. 
We note that many events have an SNR less than 20 (vertical dashed line in the figure), which is considered the detection threshold for EMRIs.

The number of EMRIs having an $\rm SNR>20$ is 42 for the case $e_{\rm out}=0.1$ and 27 for $e_{\rm out}=0.7$, corresponding to the $\sim 15\%$ and $\sim 10\%$ of all events, respectively. This means, on average, $\sim 7-10$ detections per year of EMRIs triggered by a MBHB. 
This detection rate is almost two orders of magnitude greater than $0.12 \rm yr^{-1}$, i.e. the value estimated in \citet{BodeWegg}. 

Comparing these numbers to the those obtained by \citet{PhysRevD.95.103012} for the standard formation channel, we see that MBHBs may contribute a non negligible number of the total EMRI rate. Indeed in \citet{PhysRevD.95.103012} the authors computed for their fiducial model (M1 in their article) a number of EMRI detections $\sim 189\ \rm yr^{-1}$ with the AK model developed for a Schwarzschild primary MBH. Taking the same EMRI catalogue of \citet{PhysRevD.95.103012}, but with a simplified version of the AK model, \citet{Bonetti2020} showed that the expected number of EMRIs from the two-body scenario observed during the 4 yr of the LISA mission is $\sim 370$, namely $\sim 93\ \rm yr^{-1}$. This second rate is a factor of two smaller with respect to the rate obtained in \citet{PhysRevD.95.103012}, and this is probably due to the fact that in \citet{PhysRevD.95.103012} the SNR of each EMRI is computed taking into account the inclination and polarization of each EMRI and not the sky-inclination-polarization--averaged fluxes as in \citet{Bonetti2020}.
Since we used the same code of \citet{Bonetti2020}, for a comparison we refer to the detection rates presented in that work. Thus, we can say that $\sim 10-15\%$ of the global number of EMRIs that are estimated to be detected by LISA might be triggered by MBHBs.

From Fig.~\ref{fig:SNR} we also note that there are some EMRIs having an SNR$\sim 10^2-10^3$, which would be very loud sources, extremely significant for extracting the EMRI physical parameters, as we discussed in Section~\ref{sec:Lisa rate 1}.

\begin{figure}
    \centering
    \includegraphics[width=1\linewidth]{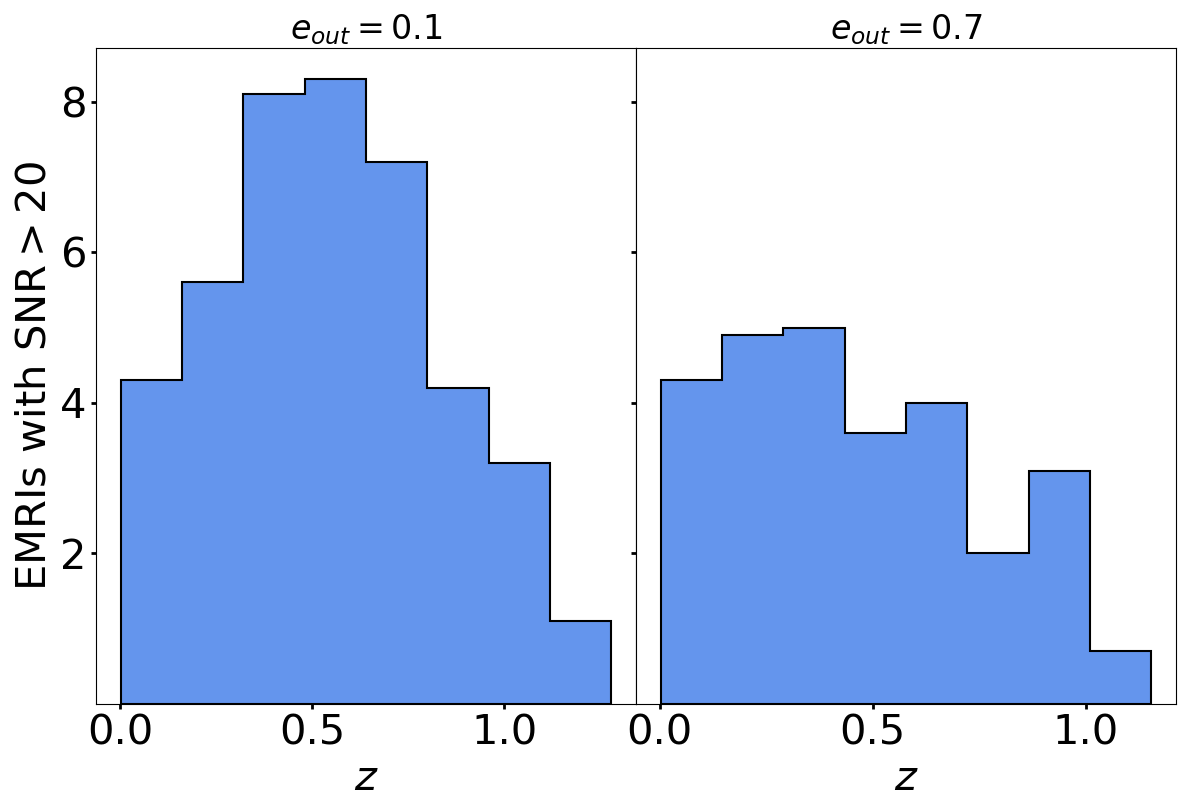}
    \caption{Number of EMRIs originated in a MBHB with eccentricity $e_{\rm out}=0.1$ (left) and $e_{\rm out}=0.7$ (right) as a function of $z$. Only sources with $\rm SNR>20$ are considered.}
    \label{fig:SNR_z}
\end{figure}

\begin{figure}
    \centering
    \includegraphics[width=1\linewidth]{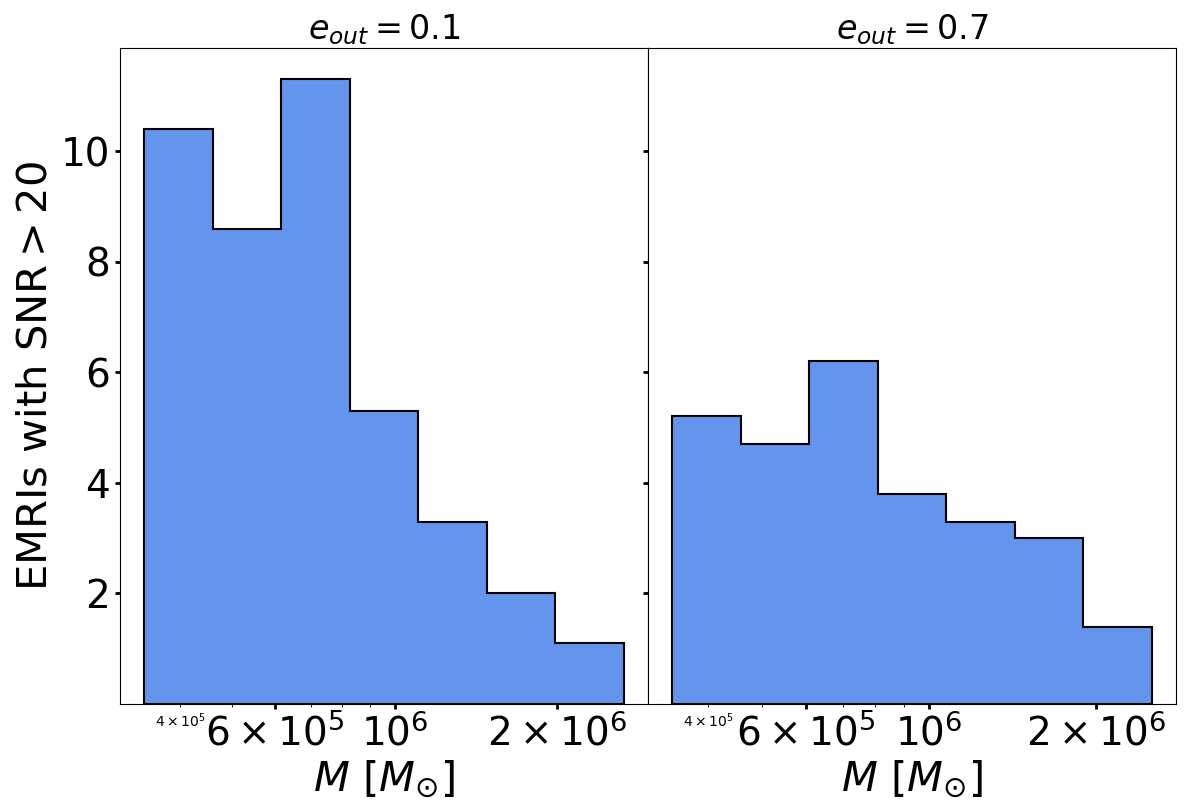}
    \caption{Number of EMRIs originated in a MBHB with eccentricity $e_{\rm out}=0.1$ (left) and $e_{\rm out}=0.7$ (right) as a function of the primary's mass. Only sources with $\rm SNR>20$ are considered.}
    \label{fig:SNR_M}
\end{figure}

Looking at the redshift distribution of the events having an $\rm SNR>20$ (Fig.~\ref{fig:SNR_z}) we see that the majority of the resolved EMRIs originate from $z<1.2$. 
Looking instead at the distribution of the primary's mass of detected EMRIs (showed in Fig.~\ref{fig:SNR_M}), we observe that it is peaked around a mass $\sim 6\times 10^5 M_{\odot}$ both for $e_{\rm out}=0.7$ and $e_{\rm out}=0.1$. These results are in contrast with the EMRI rate distribution over $M_1$ from Fig.~\ref{fig:fr}, where the EMRI formation rate monotonically increases with the mass of the primary MBH and is a direct consequence of the LISA sensitivity curve that preferentially selects lighter MBHs.
In fact, EMRIs made of a BH with mass of $10 M_{\odot}$ orbiting a $3\times 10^5 M_{\odot}$ MBH have a GW emission right in the center of the sweet spot of the LISA frequency range. If we consider a ten-time-heavier MBH, EMRI emission would be one order of magnitude shifted towards low frequencies, and thus no longer in the center of the LISA sensitivity band. Consequently, although the number of EMRIs generated by MBHs with $M>3\times 10^6 M_{\odot}$ is greater, their detectability with LISA would be much smaller. On the contrary, though Fig.~\ref{fig:fr} shows that MBHs below $6\times 10^5 M_{\odot}$
give a minor contribution to the global EMRI rate, those can still be detectable by LISA. Therefore, extending our simulations to lower masses of the primary might result in a MBHB-driven EMRI rate as much as 50\% higher.

\section{Discussion and conclusion}\label{sec:conclusions}

In this paper we explored the process of EMRI formation in MBHBs, expanding and updating the original investigations of \cite{BodeWegg} (BW14 hereafter). Our findings leverage on three main aspects: the careful assessment of the dynamics leading to EMRI formation when a secondary MBH is introduced, the computation of the EMRI cosmological formation rate based on the abundance of MBHB mergers, and finally the forecast of the number of EMRI detections achievable by LISA.

To this end, we performed $4.8\times 10^5$ three-body simulations of a stellar BH of $10M_{\odot}$ orbiting a primary MBH $M_1$, with a stellar potential centered on it, plus a secondary MBH $M_2$ forming an initially wider binary together with $M_1$. We considered three different masses for the primary MBH, $M_1\in[3\times 10^5, 10^6, 3\times 10^6]M_{\odot}$, four values of the mass-ratio between the two MBHs $q\in[0.003,0.01,0.03,0.1]$, and two different eccentricities of the MBHB, namely $0.1$ and $0.7$.
From the full sample of simulations we found that the number of EMRIs generally (see Tab.~\ref{tab:risultati_sim}): (i) increases by decreasing the mass of the primary MBH; (ii) increases by decreasing the mass-ratio of the MBHB; (iii) is higher for more circular MBHBs when $M_1 \lesssim 10^6 M_{\odot}$ and $q$ is above 0.03, while the opposite is true for larger $M_1$.

By analyzing the EMRI features we deduced that both secular (mainly LK oscillations) and chaotic interactions can drive the EMRI formation. In particular the heavier the secondary MBH and the more circular the MBHB are, the larger is the number of EMRIs originated by secular interactions; on the contrary, for lighter secondary MBHs and for $e_{\rm out}=0.7$ chaotic interactions are found to dominate EMRI production.

When assuming that stellar BHs are distributed following a SIS profile around the primary MBH, we found that the EMRI formation channel involving a MBHB leads to an EMRI rate of the order of $\sim 10^{-6}-10^{-5}\ \rm yr^{-1} $, a factor $10-100$ larger than the estimate for the standard two-body relaxation formation channel. Though seemingly large, such rate is not steady-state as the standard two-body relaxation formation rate, since in the case at hand EMRI production is limited by the lifetime of the MBHB. On average, we have a burst in EMRI formation after $10^7$ yr from the MBHB formation, then the process is halted by continuous ejection of the BHs until the final coalescence of the MBHB. 

We then evaluate the cosmic MBHB merger rate through a semi-analytical model of galaxy formation and evolution, \textit{L-Galaxies}, that provides us with a list of galaxy mergers along the cosmic history. By combining the number of EMRIs -- obtained varying the mass of the primary MBH $M_1$ and the mass-ratio $q$ -- with the MBHB formation rate we estimated a cosmological EMRI formation rate of $\sim 74 \rm \ yr^{-1}$ when $e_{\rm out}=0.1$, and of $\sim 64 \rm \ yr^{-1}$ when $e_{\rm out}=0.7$.

Finally, we assessed how many EMRIs could be potentially detected by the future GW detector LISA, i.e. those sources that exceed the SNR threshold of $\sim 20$. 
To this aim, we extracted a population of individual EMRIs by performing a Monte Carlo sampling of the cosmological EMRI rate distribution (Eq.~\eqref{eq:rate_tot}) over $M_1$, $q$ and $z$.
For each event, the LISA SNR is evaluated using the waveform model and detector assumptions described in \cite{Bonetti2020}. 
This procedure yields ${\cal O}(10)$ MBHB-induced EMRIs per year, which we estimate to be around $\sim 10\% -15 \%$ of the total number of EMRIs that LISA will detect. The latter is estimated from the results presented in \citet{Bonetti2020} and based on the fiducial model of \citet{PhysRevD.95.103012}. 

\subsection{Comparison with previous works}
\label{sec:previous_works}

BW14 analysed the same EMRI formation channel studied in this work, but following a different approach and covering a narrower parameter space.
The authors performed four groups of simulations considering two possible values of the mass-ratio of the MBHB, $q=0.1$ and $q=0.3$, and two masses for the stellar mass object, $1 \rm M_{\odot}$ (i.e. a proxy for white dwarfs or neutron stars) or $10 \rm M_{\odot}$ (i.e. a stellar BH). The mass of the primary MBH is always set equal to $M_1=10^6 M_{\odot}$. 
Their simulations feature a stalled MBHB, i.e. no orbital evolution of the secondary MBH is involved, and describe the mutual gravitational interaction via the introduction of a pseudo-Newtonian potential for the MBHs which accounts for the GR conservative effects, but no live GW dissipation is included. The effect of GW emission is accounted for only at pericentre passage of the COs and each system is integrated at most up to 1.5 Myr.

They obtained a MBHB-driven EMRI formation rate, considering only stellar BHs as EMRI progenitors, equal to $\mathcal{R}_{\rm EMRI}=8\times10^{-4}\ \rm yr^{-1} Gpc^{-3}$ for a MBHB's mass-ratio $q=0.1$ and equal to $\mathcal{R}_{\rm EMRI}=6\times10^{-4}\ \rm yr^{-1} Gpc^{-3}$ for a mass-ratio $q=0.3$.
The estimated LISA detection rate is assessed assuming full ($100\%$) detectability for EMRIs and that the EMRI GW signal is detectable only for $z<1$. The final estimated rate is $\sim 0.12 \ \rm yr^{-1}$.

For comparison here we obtain a cosmological EMRI formation rate of $R_{\rm EMRI}=74 \ \rm yr^{-1}$ for the case $e_{\rm out}=0.1$ and $R_{\rm EMRI}=64 \ \rm yr^{-1}$ for the case $e_{\rm out}=0.7$. In order to compare these results with those obtained in BW14, we can consider the global EMRI formation rate only for $z<1$, then dividing by the volume of the observable universe at that redshift, namely $160 \ \rm Gpc^{3}$. Averaging between the two values of $e_{\rm out}$, we obtain a rate of $\mathcal{R}_{\rm EMRI}= 0.14 \ \rm yr^{-1} Gpc^{-3}$, namely $\sim 200$ times greater. Note that this rate is $\approx1/7$ of the global EMRI formation rate estimated for the standard formation channel by \citet{Gair_barack_2004}, where they computed a value of $\mathcal{R}_{\rm EMRI} \sim 1 \ \rm yr^{-1} Gpc^{-3}$.

There might be several reasons for the discrepancy among the results presented above. First of all, we used different numerical methods both for the simulations, the computation of the MBHB merger rate, and the GW detectability.
In particular, for the simulations we used a three-body integrator featuring GR corrections introduced through the PN approach, instead of a Pseudo-Newtonian potential like in BW14. Moreover, the dynamical evolution of the MBHB is captured by introducing an additional dissipative force that determines a shrinkage of the semi-major axis due to the stellar hardening process. This feature has important consequences on the dynamics of the system and might be the main source of difference with respect to BW14. 
Indeed, for each simulation, BW14 started with a set-up in which there were $10^{6}$ non interacting stars, and following their evolution they obtained only a dozen of EMRIs per simulation. Moreover, also their choice to stop the simulations after $\sim 1.5 \rm Myr$, in order to limit the computational time, may have led to missing the identification of a significant fraction of EMRIs that might occur at later times.  
In our simulations we also defined different conditions with respect to BW14 for the identification of an EMRI or a DP. 

For the cosmological EMRI rate estimation, we computed first the MBHB merger rate from the output of $L-Galaxies$, a SAM of galaxy formation, rather than considering the simplistic assumption of one merger per MBH lifetime as chosen by BW14. Feeding the information about the MBHB merger rate we are able to compute the cosmological EMRI formation rate as a function of the primary's mass $M_1$, the mass-ratio $q$ of the MBHB, and the redshift $z$.
Finally, the estimation of GW signal detection is performed through a more realistic, though approximate, waveform model \citep{Bonetti2020}.

While finalising the present work, we became aware of the paper by \citet{Naoz2022}, focused on the very same problem of EMRI formation driven by the interaction with a MBHB, despite employing different techniques.
Specifically, those authors probe the EMRI formation through a secular approach, in which orbital-averaged equations of motion are numerically solved to explore the dynamics of stellar-mass BHs orbiting a $10^7 \rm M_{\odot}$ MBH and perturbed by a $10^9\rm M_{\odot}$ secondary. In addition to Newtonian mechanics, they considered 1 PN GR effects, responsible for pericentre precession, the dissipative action of GWs, as well as a stochastic description of two-body relaxation, through which a BH gets randomly kicked over a timescale of the order of the orbital period.
Besides considering a CO bound to the secondary MBH and perturbation from a more massive primary, as opposed to our work they also do not consider an evolving MBHB, which in our case is responsible for several chaotic encounters. This is a feature that cannot be captured by a secular formalism since the triplet is no more in an hierarchical configuration. Overall this could explain the high fraction of EMRIs \citet{Naoz2022} obtained given the fixed hierarchical configuration, LK oscillations have enough time to develop and sensibly increase the eccentricity of the EMRI progenitors, with the effect of relaxation bringing additional systems in an optimal configuration for LK mechanism to operate. In fact, they stress that the combination LK oscillations and relaxation can significantly increase the EMRI formation rate in this channel. While this is true, two body relaxation also scatters EMRI formation over much longer timescales (their Fig.~4), and orbital evolution of the binary could start to play a significant role. Inclusion of random kicks in our direct-integration code should be straightforward and we plan to check the effects of relaxation-induced stochasticity of the BHs' orbits in our configuration in a future work.

As for their total rates, shown in their Fig.~5, it is unclear for how long the KL enhancement can be sustained. Whether this channel can contribute significantly to the total EMRI rates will ultimately depend on the cosmic merger rate of MBHBs and on their hardening timescales and is certainly a topic that deserves further investigation.

\subsection{Caveats and outlook}
\label{sec:caveats}

We conclude by cautioning that our work is subjected to a number of assumptions and caveats, which in some cases are used to simplify the problem under consideration and in others genuinely reflects our limited knowledge of the astrophysics and dynamics of dynamical processes in galactic nuclei and EMRI formation.

In particular, it should be noted that when computing the rates, we scaled the EMRI distribution according to a SIS, which is different than the Hernquist extended stellar component employed to introduce precession in the simulations. The assumption of a SIS-like profile in the inner region is supported by long term evolution studies of dense stellar systems \citep{2006ApJ...649...91F}, where massive objects are found to form a steep cusp towards the center due to mass segregation, pushing the lighter stellar component outwards. For a fully consistent simulation, one should implement an external potential consistent with the two-component total density profile, which requires further modification and testing to the code. We plan to include this modification in future work, but we briefly justify here the impact of our choices on the outcome of this study. Employing a steeper density profile would result in shorter Newtonian precession timescales that might hamper the onset of LK oscillations. This, in practice corresponds to shifting downward the red points in Fig.~\ref{fig:Timescales}. Even a downward shift of two orders of magnitude (resulting from an extreme increase of the central density of the same factor), would only bring the tails of the red points below the LK line, mostly for systems with $q\leq 0.03$. This means that a shorter precession timescale can hamper LK oscillations only for wide inner binaries when the outer binary has a small mass ratio. In these cases, however, LK oscillations are already sub-dominant in driving EMRI formation, being the vast majority of EMRIs produced by chaotic interactions (cf, Fig.~\ref{fig:3e5_3e-3_1e-1_E} and Fig.~\ref{fig:chaos}). The marginal importance of Newtonian precession is also corroborated by the tests performed in Appendix~\ref{sec:appA}, where we found no significant difference in EMRI rates when removing the stellar potential. Because of these considerations, we are confident that the results presented in this work are robust.

Another limiting factor related to the introduction of an analytical stellar potential, is that it cannot capture small stochastic perturbations due to close encounters between the EMRI candidate and other stars and COs. It has been shown by  \citep[see e.g.][]{Naoz2022} that such perturbations can have a significant impact in the EMRI formation process. In future work, we plan to explore the effect of stochastic perturbations by implementing periodic random kicks to the velocity of the CO during the integration of the triple system.

Concerning the small-scale dynamics, in this work we considered only Schwarzschild MBHs. When spin is included, the last stable orbit of an object orbiting a MBH changes, which can have an important impact on EMRI rates, as reported in \citet{Seoane2013}. Moreover, the GW signal for a prograde EMRI around a spinning MBH is stronger, which might result in more events observable by LISA. 

The overall cosmic rates have been calculated on the bases of the MBHB merger rate predicted by $\textit{L-Galaxies}$, a semi-analytical cosmological model constructed on top of the Millenium II simulation. MBHB merger rate estimates have a significant scatter based on the underlying model assumptions, which will be reflected in the number of MBHB-triggered EMRIs. Moreover, we stress that the $\textit{L-Galaxies}$ implementation used here assumes instantaneous MBHB coalescence following galaxy mergers. One might expect that the inclusion of detailed MBHB dynamics in $L-Galaxies$ could have two competing consequences. The first is that not all the MBH pairs would merge, the second is that the MBHB-driven EMRI formation peak might shift at lower $z$, promoting their detection.

When it comes to LISA rates, those were computed using a simplified version of the analytic kludge model constructed by \citet{Barak_cutler_2004}. 
Perhaps more importantly, we considered the sky-inclination-polarization--averaged GW signals, which is conservative in terms of SNR. Since we consider only non-rotating MBHs, the EMRI waveform is also truncated at the Schwarzschild innermost stable circular orbit (ISCO).

Finally, we sampled a fairly restricted region of the parameter space in terms of mass and mass-ratio of the MBHB. As shown in Fig.~\ref{fig:SNR_M}, the simulated mass range appears to capture the peak of LISA EMRIs, missing however a significant fraction of systems at the low mass end. Moreover, MBHBs with $q>0.1$ can still effectively trigger EMRIs, even though the strong perturbation of the secondary is likely to enhance the fraction of ejections \citep{2011ApJ...729...13C}. The combination of both effects might result in a factor $\approx 2$ underestimation of the MBHB-triggered EMRI rates, which, in fact, can contribute a significant fraction of the overall EMRIs observable by LISA. Along the same lines, an additional possible underestimation of the EMRI rate could derive from our choice of neglecting COs around the secondary MBH \citep[see e.g.][who, because of the large MBH masses involved, consider the EMRI formation only for the secondary MBH]{Naoz2022}. However, we expect the number of COs bound to a given MBH to be, in first approximation, proportional to its mass and since we considered light secondaries (with masses in the range $M_2\in[~10^3,~10^5]M_{\odot}$), the overall contribution to the EMRI rate, scaling with the secondary mass, is very likely subdominant and can be safely neglected.

In future work, we plan to expand the domain in our investigation and to relax some of the aforementioned caveats to produce a more reliable rate estimate. Moreover, we plan to investigate whether MBHB-triggered EMRIs have distinctive properties that will allow to separate them from those produced by the standard two-body relaxation route. This would provide a novel route to study the population of inspiralling MBHBs without necessarily relying on electromagnetic identification. 

\section*{Acknowledgements}

A.S. acknowledges financial support provided under the European Union’s H2020 ERC Consolidator Grant "Binary Massive Black Hole Astrophysics" (B Massive, Grant Agreement: 818691). M.B. acknowledges the CINECA award under the ISCRA initiative, for the availability of high-performance computing resources and support. Numerical calculations have also been made possible through a CINECA-INFN agreement, providing access to resources on GALILEO and MARCONI at CINECA.  

\section*{Data Availability}

The data underlying this article will be shared on reasonable request to the corresponding author.



\bibliographystyle{mnras}
\bibliography{example} 

\appendix

\section{Centre of stellar potential}
\label{sec:appA}

In the original version of the code for the integration of the three-body system the stellar potential was centered in the origin of the reference frame, chosen to be the located at the center of mass of the stalled MBHB \citep[see][for details]{Bonetti1}. Conversely, in this work we decided to center it on $M_1$, to let it follow the motion of $M_1$ during the evolution of the triplet. This is because we are mostly interested in the dynamics on scales where EMRIs might be produced, i.e. well within the sphere of influence of $M_1$ (mostly  within $10^{-2}-10^{-5}$ pc). At such small distances the potential is dominated by the MBH and also the central cusp will follow the MBH in its motion, rather than remaining fixed around a specific position (i.e. the origin of the reference frame).
\begin{figure}
    \centering
    \includegraphics[width=1\columnwidth]{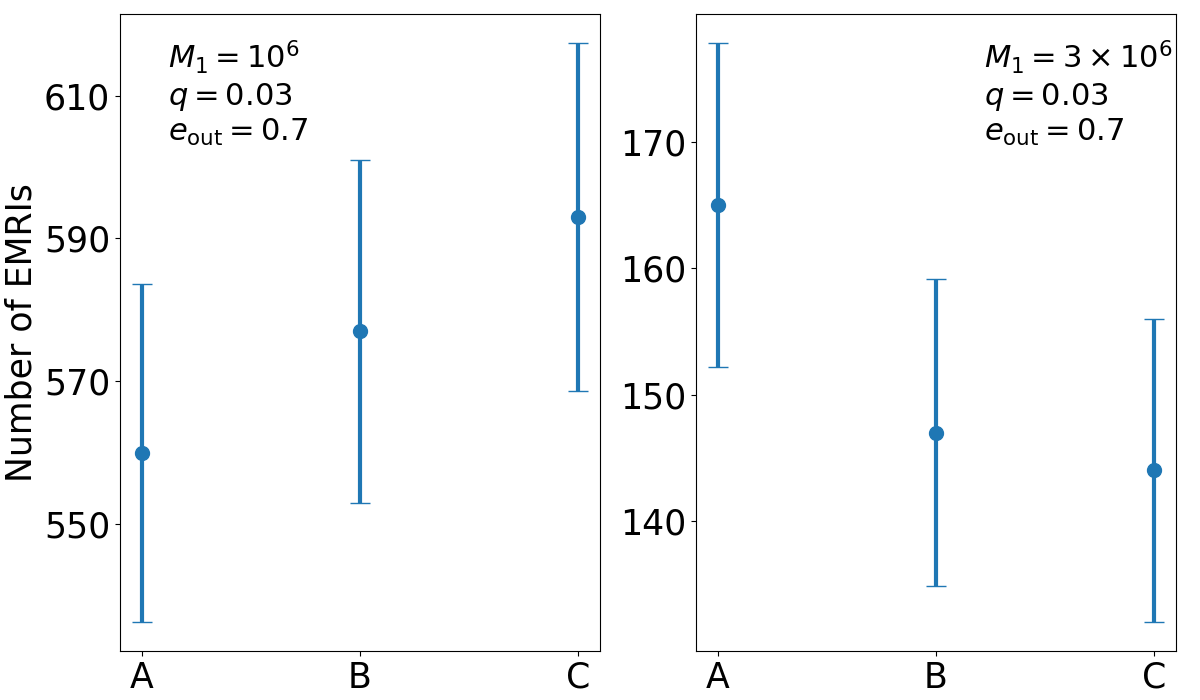}
    \caption{Number of EMRIs generated in the case $M_1=10^6 M_{\rm \odot}, \ q=0.03,\ e_{\rm out}=0.1$ (left) and $M_1=3 \times 10^6 M_{\rm \odot}, \ q=0.03,\ e_{\rm out}=0.7$ (right), with respect to the three different potential choices: A represents the case where the stellar potential is kept fixed at the origin, B considers the potential attached to $M_1$ and C neglects the stellar potential altogether. Error bars are evaluated from Poisson noise.}
    \label{fig:pot}
\end{figure}
In order to estimate the consequences of this choice on EMRI formation, we tested the effects of the stellar potential considering three different cases: the stellar potential is fixed in the center of the reference frame (A); the stellar potential centered on the primary MBH and free to move with it during the evolution of the system (B); and no stellar potential (C).
We counted the number of EMRIs in the three configurations considering 20000 simulations for each of the two following cases: $M_1=10^6 M_{\rm \odot}, \ q=0.03,\ e_{\rm out}=0.1$ and $M_1=3 \times 10^6 M_{\rm \odot}, \ q=0.03,\ e_{\rm out}=0.7$. The results are reported in Fig.~\ref{fig:pot}. The number of identified EMRIs for the three configurations and for both the cases lies between the Poissonian error bars calculated as $N\pm \sqrt{N}$. We therefore conclude that the choice of moving the origin of the stellar potential from the center of the reference frame to the position of $M_1$ does not significantly alter the EMRI formation rate, showing that the Newtonian precession both of the secondary MBH and of the stellar BH does not play an important role in the formation of EMRIs through this channel.

\bsp	
\label{lastpage}

\end{document}